\documentclass[prd,aps,floats,epsfig,eqsecnum,nofootinbib]{revtex4}
\usepackage{amsmath,amssymb,verbatim,epsfig}
\begin{document}
\title{Primordial magnetic fields from cosmological phase transitions}
\footnote{To appear in the Proceedings of the 9th. International
  School of Astrophysics `Daniel Chalonge', N. Sanchez and
  Y. N. Parijskij Editors, NATO ASI Series, Kluwer Publ.}
\author{D. Boyanovsky$^{(a,b)}$}
\email{boyan@pitt.edu}
\author{H. J. de Vega$^{(b,a)}$}
\email{devega@lpthe.jussieu.fr}
\author{M. Simionato $^{(a)}$}
\email{mis6@pitt.edu} \affiliation{$^{(a)}$ Department of Physics
and Astronomy,
University of Pittsburgh, Pittsburgh, Pennsylvania 15260, USA\\
$^{(b)}$ LPTHE, Universit\'e Pierre et Marie Curie (Paris VI) et
Denis Diderot (Paris VII), Tour 16, 1er. \'etage, 4, Place
Jussieu, 75252 Paris, Cedex 05, France}

\begin{abstract}
We study the generation of large scale primordial magnetic fields
by a cosmological phase transition during the radiation dominated
era. The setting is a theory of $N$ charged scalar fields coupled
to an abelian gauge field, that undergoes a phase transition at a
critical temperature much larger than the electroweak scale. The
dynamics after the transition features two distinct stages: a
spinodal regime dominated by linear long-wavelength instabilities,
and a scaling stage in which the non-linearities and backreaction
of the scalar fields are dominant. This second stage describes the
growth of horizon sized domains. We implement a formulation based
on the non-equilibrium Schwinger-Dyson equations to obtain the
spectrum of magnetic fields that includes the dissipative effects
of the plasma. We find that large scale magnetogenesis is
efficient during the scaling regime. Charged scalar field
fluctuations with wavelengths of the order of the Hubble radius
induce large scale magnetogenesis via \emph{loop effects}. The
leading processes are: pair production, pair annihilation
and low energy bremsstrahlung, these processes while forbidden in
equilibrium are allowed strongly out of equilibrium. The ratio
between the energy density on scales larger than $L$ and that in
the background radiation $r(L,T)= \rho_B(L,T)/\rho_{cmb}(T)$ is
$r(L,T) \sim 10^{-34}$ at the Electroweak scale and $r(L,T)\sim
10^{-14}$ at the QCD scale for $L \sim 1~\mbox{Mpc}$. The
resulting spectrum is insensitive to the magnetic diffusion length
and equipartition between electric and magnetic fields does
{\bf not} hold. We conjecture that a similar mechanism could be
operative after the QCD chiral phase transition.
\end{abstract}
\date{\today}

\maketitle

\tableofcontents 

\section{Introduction}
A variety of astrophysical observations including Zeeman
splitting, synchrotron emission, Faraday rotation measurements
(RM) combined with pulsar dispersion measurements (DM) and
polarization measurements suggest the presence of large scale
magnetic fields\cite{Parker,Kronberg,grasso,widrow,giova1,han}.
The  strength of typical galactic magnetic fields is measured to
be $\sim \mu~G$\cite{Kronberg,grasso,widrow,han} and they are
correlated on very large scales up to galactic or even larger
reaching to scales of cluster of galaxies $\sim
1~\mbox{Mpc}$\cite{Kronberg,grasso,widrow,giova1}. The origin of
these large scale magnetic fields is still a subject of much
discussion and controversy. It is currently agreed that a variety
of dynamo mechanisms are efficient in {\bf amplifying} seed
magnetic fields with typical growth rates $\Gamma \sim
\mbox{Gyr}^{-1}$ over time scales $\sim 10-12 $ Gyr (for a
thorough discussion of the mechanisms and  models
see\cite{widrow}). The ratio of the energy density of the seed
magnetic fields  on scales larger than $L$ (today) to that in the
cosmic background radiation, $r(L)=\rho_B(L)/\rho_{cmb}$ must be
$r(L\sim 1\mbox{Mpc})\geq 10^{-34}$ for a dynamo mechanism to
amplify it to the observed value, or $r(L\sim 1\mbox{Mpc})\geq
10^{-8}$ for the seed to be amplified solely by the gravitational
collapse of a protogalaxy\cite{giova1}.

There are also different  proposals to explain the origin of the
initial seed. Astrophysical batteries rely on gradients of the
charge density concentration and pressure and their efficiency in
producing seeds of the necessary amplitude is still very much
discussed\cite{Kronberg,widrow}. Primordial magnetic fields that
could be the seeds for dynamo amplification can be generated at
different stages in the history of the early Universe, in
particular during inflation, preheating and or phase
transitions\cite{grasso,giova1,widrow}. Primordial (hyper)
magnetic fields may have important consequences in electroweak
baryogenesis\cite{grassoEW}, Big Bang nucleosynthesis
(see\cite{grasso}), the polarization of the CMB\cite{durrer1} via
the same physical processes as Faraday rotation, and structure
formation\cite{grasso,giova1,dolgov2}, thus sparking an intense
program to study the origin and consequences of the generation of
magnetic fields in the early Universe\cite{hogan}-\cite{ahonen}.

A reliable estimate of the amplitude and correlations of seed
magnetic fields must include the dissipative properties of the
plasma, in particular the
conductivity\cite{turnerwidrow,Giovanni,giovashapo}. In
ref.\cite{magfiI} we have  introduced a formulation that allows to
compute the generation of magnetic fields from processes strongly
out of equilibrium. This formulation, which is based on the exact
set of Schwinger-Dyson equations for the transverse photon
propagator is manifestly gauge invariant and is general for any
matter fields and any cosmological background (conformally related
to Minkowski space-time). In the case in which strongly out of
equilibrium effects arise from long-wavelength fluctuations, such
as during phase transitions, this formulation allows to separate
the contribution of the hard degrees of freedom which are in local
thermodynamic equilibrium from that of the soft degrees of freedom
that fall out of LTE (local thermal equilibrium) during the phase
transition and whose dynamics is strongly out of equilibrium. This
separation of degrees of freedom leads to a consistent
incorporation of the dissipative effects via the conductivity (for
details see\cite{magfiI}). In that reference a  study of
magnetogenesis in Minkowski space-time during a supercooled phase
transitions was presented and the results highlighted the main
aspects of the generation of magnetic and electric fields in these
situations.

We study the generation of large scale (hyper) magnetic fields by
a cosmological phase transition during a radiation dominated era
by implementing the formulation introduced in ref.\cite{magfiI}.
The setting is a theory of $N$ charged scalar fields coupled to an
abelian gauge field (hypercharge). We consider the situation when
this theory undergoes a phase transition after the reheating stage
and before either the Electroweak or the QCD phase transition,
since we expect that these transitions will lead to new physical
phenomena. The non-perturbative dynamics out of equilibrium is
studied in the limit of a large number $N$ of (hyper) charged
fields and to leading order in the gauge coupling. The
non-equilibrium dynamics of the charged scalar sector features two
distinct stages. The first one describes the early and
intermediate time regime and is dominated by the spinodal
instabilities which are the hallmark of the process of phase
separation and domain formation and growth. This stage describes
the dynamics between the time at which the phase transition takes
place and that at which non-linearities become important via the
backreaction. The second stage corresponds to a \emph{scaling
regime} which describes the slower non-equilibrium evolution of
Goldstone bosons and the process of phase ordering\cite{scaling}
and growth of horizon-sized domains. This scaling regime is akin
to the solution found in the \emph{classical} evolution of scalar
field models with broken continuous symmetries after the phase
transition that form the basis for models of structure formation
based on topological defects\cite{turok,durrer}.

The solution of the scalar field dynamics \cite{scaling} is the
input in the expression for the spectrum of the magnetic field
obtained in~\cite{magfiI} to obtain the amplitude of the
primordial seed generated during both stages.

We find that scaling stage is the most important  for the
generation of large scale magnetic fields. Large scale magnetic
fields are generated via loop effects from the dynamics of modes
that are at the scale of the horizon or smaller. The leading order
processes that result in the generation of large scale magnetic
fields are: i) pair {\bf production}, ii) pair annihilation and
iii) low energy bremsstrahlung. These processes  would be
forbidden in equilibrium by energy momentum conservation, but they are
allowed strongly out of equilibrium because of the rapid time
evolution of the cosmological background and the fast dynamics of the
scalar field fluctuations. 

The resulting spectrum is rather insensitive to the diffusion
length scale which is much smaller than the horizon during the
radiation dominated era. The ratio of the magnetic energy density
on scales larger than $L$ (today) to the energy density in the
background radiation $r(L,\eta)=\rho_B(L,\eta)/\rho_{cmb}(\eta)$
is summarized in a compact formula [eq.(\ref{rreges})]. For $L
\sim 1~\mbox{Mpc}$ (today) we find $r(L,\eta) \sim 10^{-34}$ at
the Electroweak scale and $r(L,\eta)\sim 10^{-14} $  at the QCD
scale, suggesting the possibility that these primordial seeds
could be amplified by dynamo mechanisms to the values of the
magnetic fields consistent with the observed ones on these scales.

  \section{The physical picture}
Whether a symmetry breaking phase transition occurs in local
thermodynamic equilibrium (LTE) or not is a detailed question of
time scales. Two important time scales must be compared, the time
scale of cooling through the critical temperature
$\tau_{cool}=T(\eta)/\dot{T}(\eta)$ and the relaxation time scale
of a fluctuation of typical wavelength $k$, $\tau(k)$. If cooling
through $T_c$  occurs on time scales \emph{slower} than the
relaxation scale, i.e, $\tau_{cool} \gg \tau(k)$ then a
fluctuation of wavevector $k$ adjusts to LTE and for these scales
the phase transition occurs in LTE. On the other hand if
$\tau_{cool} \ll \tau(k)$ fluctuations of wavelength $k^{-1}$
cannot adjust to the conditions of LTE and fall out of
equilibrium, i.e, \emph{freeze out} and for these scales the phase
transition occurs on very short time scales like a quench from a
high temperature phase into a low temperature
phase~\cite{boysinglee}. While short wavelength modes typically
have very short relaxation time scales and remain in LTE during
the phase transition, long-wavelength modes undergo critical
slowing down near $T_c$ and their relaxational dynamics becomes
very slow\cite{critslowdown}. Thus as the temperature nears the
critical with a non-vanishing cooling rate, long-wavelength modes
freeze out falling out of LTE, namely  for these modes  the phase
transition occurs out of equilibrium. In theories in which the
\emph{equilibrium} phase transition is of second order, the
\emph{non-equilibrium } dynamics below the critical temperature is
described by the process of \emph{spinodal decomposition}:
long-wavelength fluctuations become unstable and grow. In
Minkowski space-time this growth is  exponential in time but in an
expanding cosmology the growth will depend on the time dependence
of the scale factor.

The growth of long-wavelength fluctuations entails that the field
becomes correlated within regions characterized by a
time-dependent correlation length $\xi(\eta)$ and the amplitude of
the long-wavelength fluctuations becomes non-perturbatively large.
The non-perturbative amplitudes of long-wavelength fluctuations,
becoming of $\mathcal{O}(1/\sqrt{\lambda})$ with $\lambda$ the
scalar self-coupling are a consequence of the fact that the
mean-square root fluctuation of the field
 probes the broken symmetry states with $\langle
\phi^2 \rangle \propto 1/\lambda$ with $\lambda$.

This is the hallmark of the process of phase separation, the
correlated regions correspond to domains, inside these domains the
field is near one of the vacuum states. A particular time scale,
the nonlinear time $t_{nl} \propto \ln(1/\lambda)$ , determines
the transition from a regime of linear instabilities to one in
which the full non-linearities become important. This time scale
roughly corresponds to when the mean square root fluctuation of
the scalar field samples the broken symmetry states and the phase
transition is almost complete. At this stage the amplitude of the
long-wavelength fluctuations become of $\mathcal{O}(1/{\lambda})$,
the phases freeze out and the long-wavelength fluctuations become
\emph{classical} but stochastic\cite{nuestros}-\cite{destri}.

Consider now the situation where the scalar field carries an
abelian charge and is coupled to the electromagnetic gauge field.
The strong spinodal fluctuations in the scalar field will induce
fluctuations in the current-current correlation function, and
while the expectation value of the current must vanish by
translational and rotational invariance, the current and charge
correlators will have strong non-equilibrium fluctuations. These
current fluctuations will in turn generate a magnetic field with a
typical wavelength corresponding to the wavelength of the
spinodally unstable modes.

This  is the main premise of this article: the spinodal
instabilities which are the hallmark of a non-equilibrium symmetry
breaking phase transition will lead to strong charge and current
fluctuations of the charged scalar fields which in turn,  lead to
the generation of magnetic fields through the non-equilibrium
evolution.

The main ingredients that must be developed in order to understand
the generation of magnetic fields through this non-equilibrium
process are:

\begin{itemize}
\item{A consistent framework to compute the spectrum of generated
magnetic field, namely $\langle \vec{B}({\vec k},t)\cdot
\vec{B}({-\vec k},t)\rangle/V$ with $\vec{B}({\vec k},t)$ the
spatial Fourier transform of the Heisenberg magnetic field
\emph{operator} and $V$ the (comoving) volume of the system.}
\item{We anticipate that plasma effects must be included to assess
the generation and eventual decay of the magnetic fields. If there
is a large conductivity in the medium the magnetic field will
diffuse but also its generation will be hindered. This will be a
point of particular importance within the cosmological
setting\cite{giova1,Giovanni,giovashapo,grasso,turnerwidrow}. }

\item{A major challenge of \emph{any} mechanism of large scale
magnetogenesis is to generate the seed magnetic fields from
microscopic, \emph{causal} processes. An important aspect of the
results presented here is that this generation mechanism is
mediated by \emph{loop} effects and correspond to processes that
are forbidden in equilibrium but allowed strongly out of
equilibrium.  }
\end{itemize}

These ingredients will be analyzed in detail below.

\section{Magnetic fields in Friedmann-Robertson-Walker cosmology}

The cosmological setting in which we are primarily interested
corresponds to a symmetry breaking phase transition in a radiation
dominated Universe. Such phase transition is in principle
different from the electroweak one\footnote{If the electroweak
phase transition is weakly first order, nucleation will be almost
indistinguishable from spinodal decomposition and the phenomena
studied here may be of relevance.} and presumably occurs at a much
higher energy scale, such as the GUT scale $\sim
10^{15}\mathrm{Gev}$ but is assumed to be described by a particle
physics model that includes many fields with (hyper)-charge either
fermionic or bosonic. We will not attempt to study a particular
gauge theory phenomenologically motivated by some GUT scenario,
but will focus our study on a generic scalar field model in which
the scalar fields carry an abelian (hyper)charge. The simplest
realization of such model is  scalar electrodynamics with $N$
charged scalar fields $ \phi_r, \; r=1, \ldots, N$ and one neutral
scalar field $ \psi $ whose expectation value is the order
parameter associated with the phase transition. The neutral field
is not coupled to the gauge field and its acquiring an expectation
value does not break the $U(1)$ gauged symmetry. This guarantees
that the abelian gauge symmetry identified with either
hypercharger or electromagnetism is \emph{not spontaneously
broken} to describe the correct low energy sector with unbroken
$U(1)_{EM}$. We will take the neutral and the $N$ complex
(charged) fields to form a scalar multiplet under an $O(2N+1)$
isospin symmetry. The electromagnetic coupling explicitly breaks
the $O(2N+1)$ symmetry down to $SU(N)\times U(1)$. In the absence
of electromagnetic coupling as the neutral field acquires an
expectation value the isospin symmetry is spontaneously broken to
$O(2N)$.  Since by construction only the neutral field acquires a
non vanishing expectation value under the isospin symmetry
breaking the photon remains massless (it will obtain a Debye
screening mass from medium effects, see discussion below).

The action that describes this theory in a general cosmological
background  is given by
\begin{equation}\label{lagra} S= \int d^4x
\sqrt{-g}\left[g^{\mu\nu}\left(\frac{1}{2} \partial_{\mu}\psi
 \; \partial_{\nu}\psi+ \mathcal{D}_{\mu}\phi^*  \; \mathcal{D}_\nu
\phi\right) +\mu^2 \left(\frac{\psi^2}{2}+\phi^*\phi\right)
-\frac\lambda{4N} \left(\frac{\psi^2}{2}+\phi^*\phi\right)^2
-\frac{1}{4} \mathcal{F}_{\mu \nu} \; \mathcal{F}_{\alpha \beta}
\; g^{\mu \alpha} \; g^{\nu \beta}\right]
\end{equation}
\noindent  where
\begin{equation}\label{cova}\mathcal{D}_\mu =
\partial_{\mu}-ie \mathcal{A}_\mu \quad \mbox{and} \quad
\mathcal{F}_{\mu \nu} = \partial_{\mu} \mathcal{A}_{\nu} -
\partial_{\nu}\mathcal{A}_{\mu} - i e [\mathcal{A}_{\mu},
\mathcal{A}_{\nu}] \; .
\end{equation}
 and
$$
\phi^\dagger\phi=\sum_{r=1}^N\phi^\dagger_r\; \phi_r \;,\quad
D_\mu\phi^\dagger D^\mu\phi=\sum_{r=1}^N(\partial_\mu+ieA_\mu)
\phi_r^\dagger\;(\partial^\mu-ieA^\mu)\phi_r\;.
$$
Furthermore, anticipating a non-perturbative treatment of the
non-equilibrium dynamics of the scalar sector in a large $N$
expansion, we have rescaled the quartic coupling in such a way as
to display the contributions in terms of powers of $ 1/N $,
keeping $\lambda$ fixed in the large $N$ limit.

A Friedmann-Robertson-Walker line element
\begin{equation}\label{FRWds}
ds^2= dt^2-a^2(\eta) \; d{\vec x}^2 \; ,
\end{equation}
\noindent is conformally related to a Minkowski line element by
introducing the  conformal time $\eta$  and scale factor $C(\eta)$
as
\begin{equation}
\eta = \int \frac{dt}{a(\eta)}~~;~~ C(\eta) = a(\eta(\eta))\; ,
\end{equation}
In terms of these the line element and metric are given by
\begin{equation}\label{conformal}
ds^2 = C^2(\eta) \; (d\eta^2 - d{\vec x}^2)~~;~~g_{\mu \nu}=
C^2(\eta)  \; \eta_{\mu \nu}\; ,
\end{equation}
\noindent where $\eta_{\mu\nu}=\mbox{diag}(1,-1,-1,-1)$ is the
Minkowski metric. Introducing the conformal fields
$$
A_{\mu}(\eta,\vec x)=\mathcal{A}_{\mu}(t(\eta),\vec x)~;~
\Phi(\eta,\vec x) = C(\eta) \; \phi(t(\eta),\vec x)~;~
\Psi(\eta,\vec x) = C(\eta) \; \psi(t(\eta),\vec x)
$$
and in terms of the conformal time, the action now reads
\begin{equation}\label{confoS}
S= \int d\eta\; d^3x \left[\eta^{\mu
\nu}\left(\frac{1}{2}\partial_{\mu}\Psi \partial_{\nu}\Psi+
D_{\mu}\Phi^*D_{\nu}\Phi\right)-M^2(\eta)\left(\frac{\Psi^2}{2}+
\Phi^*\Phi\right)
-\frac\lambda{4N}\left(\frac{\Psi^2}{2}+\Phi^*\Phi\right)^2-\frac{1}{4}
F_{\mu \nu} \;  F_{\alpha \beta} \; \eta^{\mu \nu} \; \eta^{\alpha
\beta}\right]
\end{equation}
\noindent with
\begin{eqnarray} \label{masconf}
&&M^2(\eta) = -\mu^2 C^2(\eta)- \frac{C''(\eta)}{C(\eta)} \quad ,
\quad D_{\mu} = \partial_{\mu}-ie A_{\mu} ~~; ~~ F_{\mu \nu} =
\partial_{\mu }A_{\nu}-\partial_{\nu}A_{\mu}\; ,
\end{eqnarray}
\noindent and the primes refer to derivatives with respect to
conformal time.  Obviously the conformal rescaling of the metric
and fields turned the action into that of a charged scalar field
interacting with a gauge field in \emph{flat Minkowski
space-time}, but the scalar field acquires a time dependent mass
term\footnote{Here we neglect the effect of the conformal
anomaly\cite{dolgovanomaly}}. In particular, in the absence of
electromagnetic coupling, the equations of motion for the gauge
field $ A_{\mu} $ are those of a free field in flat space time.
This is the statement that gauge fields are \emph{conformally}
coupled to gravity and no generation of electromagnetic fields can
occur from gravitational expansion alone without coupling to other
fields or breaking the conformal invariance of the gauge sector.
The generation of electromagnetic fields must arise from a
coupling to other fields that are not conformally coupled to
gravity, or by adding extra terms in the Lagrangian that would
break the conformal invariance of the gauge
fields\cite{turnerwidrow}.

The conformal electromagnetic fields
$\vec{\mathcal{E}},\vec{\mathcal{B}}$ are related to the physical
$\vec E, \vec B$ fields by the conformal rescaling
\begin{equation}\label{physicalfields} \vec{E}=
\frac{\vec{\mathcal{E}}}{C^2(\eta)}~~;~~\vec{B}=
\frac{\vec{\mathcal{B}}}{C^2(\eta)} \; ,
\end{equation}
\noindent corresponding to fields of scaling dimension two.

\vspace{2mm}

{\bf Gauge invariance:} Since we will study the generation of
magnetic fields through loop effects and keeping only low order
terms in the diagrammatic expansion, it is important to guarantee
that the results obtained are  \emph{gauge invariant}, namely
physical.

 A gauge invariant formulation leads to the
following Lagrangian density (for details
see\cite{magfiI,gauginv})
\begin{eqnarray}\label{gauginvlag} {\cal L}= && \frac{1}{2}\partial_{\mu}\Psi\;
\partial^{\mu}\Psi+\partial_{\mu}\Phi^\dagger   \;
\partial^{\mu}\Phi+\frac{1}{2}\partial_{\mu}\vec A_T\cdot\partial^{\mu} \vec
A_T+\frac{1}{2} (\nabla A_0)^2
-M^2(\eta)\left(\frac{1}{2}\Psi^2+\Phi^\dagger\Phi\right)
-\frac\lambda{4N}
\left(\frac{1}{2}\Psi^2+\Phi^\dagger\Phi\right)^2 \nonumber \\&&
-ie\vec A_T\cdot\left(\Phi^\dagger\nabla\Phi-
\nabla\Phi^\dagger\Phi\right)- e^2(\vec
A_T^2-A_0^2)\;\Phi^\dagger\Phi-ie\; A_0\left(
\Phi\dot\Phi^\dagger-\Phi^\dagger\dot\Phi\right) \; ,
\end{eqnarray}
\noindent where $\Phi$ is a gauge invariant \emph{local} field
which is non-locally related to the original fields, and
$\vec{A}_T$ is the transverse component of the vector field
(${\vec \nabla}\cdot {\vec A}_T=0$) and  $A_0$ is a
non-propagating field as befits a Lagrange multiplier, its
dynamics is  completely determined by that of the  charge
density~\cite{magfiI,gauginv}.

The main point of this discussion is that the framework to obtain
the power spectrum of the generated magnetic field presented below
is fully \emph{gauge invariant}.

\section{Phase transitions in radiation dominated cosmology}

\subsection{Kinematics}

We consider here a phase transition in a radiation dominated
cosmology in a theory with a critical temperature near but below a
GUT scale, and we take the reheating temperature of the
post-inflationary universe to be higher than the critical
temperature. Furthermore we assume that after reheating the
Universe is described by a state of local thermodynamic
equilibrium and the background radiation dominates the dynamics of
the Hubble expansion.

In this radiation dominated cosmology, as the initial state  local
thermodynamic equilibrium at an initial temperature $T_R
>>T_c$ the expansion results in cooling and the temperature eventually
falls below the critical 
triggering the phase transition.

 Using finite temperature field
theory in an expanding background geometry, it is
shown~\cite{scaling} that the effective time dependent mass term
depends on the effective time dependent temperature
$T(\eta)=T_R/a(\eta)$ (see discussion below) which reflects the
cooling from the cosmological expansion. Hence at a given time the
temperature equals the critical and the phase transition occurs.
Field modes with wavectors much larger than the symmetry breaking
scale $\mu$ will remain in LTE and will not be affected by the
symmetry breaking dynamics\cite{boysinglee}, but those with
wavevectors much smaller than the symmetry breaking scale will
fall our of equilibrium and become spinodally
unstable\cite{boysinglee}.

We normalize the scaling factor $C(\eta)$ at the reheating time
$\eta=\eta_R$ in such a way that $C(\eta_R)=1$ then the explicit
expression for $C(\eta)$ reads
\begin{equation}\label{C-rde}
C(\eta)=H_R\;\eta
\end{equation}
\noindent where $H_R$ is the Hubble constant at the reheating
time, $H_R=\eta_R^{-1}$.

We can relate $H_R$ to the reheating temperature and the Planck
mass $ G^{-\frac12} $ through the  equation
\begin{equation}\label{blackbody}
\rho= \frac{\pi^2g_*}{30}~T_R^4
\end{equation}
and the Einstein-Friedman  equation
\begin{equation}\label{H-T}
H_R=\left(\frac83\pi \; G \; \rho\right)^{1/2}=\frac{T_R^2}{M_*}
\; ,
\end{equation} \noindent
where $g^*$ is the effective number of degrees of freedom at the
reheating temperature and we introduced the scale $M_*$ of the
order of the Planck mass
\begin{equation}\label{Mstar}
M_*=\frac{3 \; \sqrt{5}}{2 \; \pi^\frac{3}{2}}\;
\frac{1}{\sqrt{g_* \; G}} \; .
\end{equation}
In radiation dominated epoch the time-dependence of the mass term
(\ref{masconf}) is given by the expression
\begin{equation}\label{mass}
-M^2(\eta)=\mu^2 \; H_R^2 \; \eta^2=\tilde\mu^4 \; \eta^2 \; ,
\end{equation}
\noindent where we see the emergence of a new mass scale
\begin{equation}\label{mu-tilde}
\tilde\mu=\sqrt{\mu \;  H_R}\;.
\end{equation}
This scale will play an important role in the following discussion
and in the comparison with results obtained in  Minkowski
space-time\cite{magfiI}. There is a last scale which plays a
relevant role, the horizon scale $r_H(\eta)$ which is fixed by the
evolution on the time of the Hubble constant:
\begin{equation}\label{hubble}
r_H(\eta)=\frac{1}{H(\eta)}=C(\eta)~\eta=H_R~\eta^2 \; .
\end{equation}
Modes with physical wavelength
$\lambda_{phys}=\frac{2\pi}{k_{phys}}$ inside the horizon
\begin{equation}
\lambda_{phys}(\eta)\sim k_{phys}^{-1}(\eta)<r_H(\eta)
\end{equation}
are causally connected, modes outside the horizon are causally
disconnected.

The relaxation rate of hard modes of the charged fields is given
by\cite{scalarqed}
\begin{equation}\label{relrate}
\Gamma(\eta) \sim \alpha \; T(\eta) \; \ln\frac{1}{\alpha} \; ,
 \end{equation}
where the effective temperature varies with time as
\begin{equation}\label{Tef}
 T(\eta)=\frac{T_R}{C(\eta)}
\end{equation}
and  the expansion rate given by
\begin{equation}
H(\eta)=\frac{T^2(\eta)}{M_{*}} \; .
\end{equation}
Therefore,
\begin{equation} \frac{\Gamma(\eta)}{H(\eta)} \sim
\frac{10^{16}}{T(\eta)[\mbox{Gev}]} \; .
\end{equation}
Thus hard modes are in thermal equilibrium for $T(\eta)\leq
10^{15}\mbox{Gev}$.

In particular, modes with $k\sim T_R$ are the hard modes that give
the leading contribution to the conductivity in the high
temperture limit\cite{baym,yaffe}. Modes with $k < \mu$ will
manifest the long-wavelength spinodal instabilities and their
dynamics will be strongly out of
equilibrium\cite{boysinglee,nuesfrw,nuestros,destri}. Their
amplitude becomes non-perturbatively
large\cite{boysinglee,nuesfrw,nuestros,destri} and will be
responsible for the non-equilibrium generation of the primordial
magnetic field\cite{magfiI}.

Using eqs.(\ref{C-rde}) and (\ref{Tef}) we can write the conformal
time as
\begin{equation}\label{eta.T}
\eta = \frac{T_R}{H_R \; T(\eta)} = \frac{M_*}{T_R~T(\eta)}\; .
\end{equation}
As it will become clear below an important cosmological  quantity
is the product
\begin{equation}\label{k.phys}
k\;\eta=\frac{k}{C(\eta)} \; C(\eta) \;
\eta=\frac{k_{phys}(\eta)}{T(\eta)}\;T(\eta) \;
r_H(\eta)=\frac{2\pi}{LT_R}  \;   \frac{M_*}{T(\eta)} \; .
\end{equation}
The ratio
\begin{equation}
\frac{k_{phys}(\eta)}{T(\eta)}=\frac{2\pi}{L \; T_R}
\end{equation}
is a kinematical invariant. Its value today is determined by the
scale $L$ which will be typically chosen to correspond to a
galactic scale or the scale of galaxy clusters, and the
temperature of the CMB. It is given by,
\begin{equation} \label{LTR}
L \; T_R = 3.7\times 10^{25}\left(\frac{L}{\mbox{Mpc}}\right) \; .
\end{equation}
Therefore,
\begin{equation} \label{horiz}
k \; \eta \sim 10^{-9} \;
\frac{T_{EW}}{T(\eta)}\left(\frac{\mbox{Mpc}}{L} \right)=\left\{
\begin{array}{l}
  10^{-22}~ \; \mbox{for}~T(\eta)=T_R \sim 10^{15} \; \mbox{Gev} \\
  10^{-9}~ \; \mbox{for\, the\, EW\, transition} \\
  10^{-6}~ \; \mbox{for\, the\, QCD\, transition}
\end{array}\right.
\end{equation}
for $L\sim 1~\mbox{Mpc}$. Thus, during the regime of interest in
this article, $k\eta \ll 1$ for scales of galaxy clusters. A
noteworthy aspect of eq.(\ref{horiz}) is that the wavelengths
corresponding to the scale of galaxies or clusters today were well
outside the horizon during the radiation dominated era when the
electroweak and QCD phase transitions occurred.

Thus the challenge for any mechanism of primordial magnetogenesis
operating on scales well inside the Hubble radius, such as causal
microscopic processes, must generate magnetic fields with
wavelengths which are much larger than the Hubble radius.

Another important quantity is the ratio of the wavevector $k$ of
the primordial magnetic field to the conductivity.

As it will be discussed below, the physical conductivity is given
by
\begin{equation}\label{conduc}
\sigma(\eta) = \frac{\mathcal{C}  \;  N(\eta) \; T(\eta)}{\alpha
\;  \ln\frac{1}{\alpha N(\eta)}} \; ,
\end{equation}
\noindent where $\mathcal{C}$ is a constant of $\mathcal{O}(1)$,
 $N(\eta)$ is the number of ultrarelativistic charged species,
and we have neglected the (logarithmic) dependence on the energy
scale in the running coupling constant. For this discussion we
will neglect  the time dependence  of $N(\eta)$  assuming that the
number of charged ultrarelativistic species remains constant (this
assumption can be relaxed without qualitative modifications of the
main argument). Under this assumption
\begin{equation}\label{physisigma}
\sigma(\eta)= \frac{\sigma_R}{C(\eta)} \; ,
\end{equation}
\noindent with $\sigma_R$ being the \emph{comoving} conductivity
determined at the time of reheating
\begin{equation}\label{comovingsigma}
\sigma_R= \frac{\mathcal{C} \;  N \;  T_R}{\alpha
\ln\frac{1}{\alpha N}} \; .
\end{equation}
Thus the ratio,
\begin{equation}\label{otroratio}
\frac{k_{phys}(\eta)}{\sigma(\eta)}\sim
\frac{2\pi\alpha}{N\;LT_R}\sim 10^{-27} \left(\frac{\mbox{Mpc}}{L}
\right)\; ,
\end{equation}
\noindent neglecting logarithmic corrections.

Furthermore,
\begin{equation}\label{sigrat}
\sigma(\eta) \;  r_H(\eta) \sim \frac{M_*}{\alpha T(\eta)} \sim
\left\{ \begin{array}{l}
  10^{5} \; ~\mbox{for}~T(\eta)=T_R \sim 10^{15}\mbox{Gev} \\
  10^{18} \; ~\mbox{for\,the\,EW\,phase\,transition} \\
  10^{21} \; ~\mbox{for\,the\,QCD\,phase\,transition}
\end{array}\right.
 \end{equation}
\noindent where we have neglected logarithmic corrections.
Therefore $\sigma_R \; \eta \gg 1$ throughout the radiation
dominated era considered in this article. The regime $\sigma_R  \;
\eta \gg 1~;~~ k^2 \; \eta/\sigma_R \ll 1$ is dominated by the
(slow) hydrodynamic relaxation of the magnetic field.

Another relevant estimate involves the (comoving) \emph{diffusion
length} $\xi_{diff}(\eta)=\sqrt{\eta/\sigma_R}$
\begin{equation}\label{difflength}
\frac{\xi_{diff}(\eta)}{\eta} \sim \sqrt{\frac{\alpha \;
T(\eta)}{M_*}}\sim\left\{\begin{array}{l}
  10^{-3} \; ~\mbox{for}~T(\eta)=T_R \sim 10^{15}\mbox{Gev} \\
  10^{-9} \; ~\mbox{for\,the\,EW\,phase\,transition} \\
  10^{-10} \; ~\mbox{for\,the\,QCD\,phase\,transition}
\end{array}\right.
\end{equation}
\noindent where again we have neglected logarithmic terms.
Therefore the diffusion length is much smaller than the Hubble
radius during the radiation dominated era. Finally, combining
(\ref{difflength}) and (\ref{horiz}) we find
\begin{equation}\label{produ}
10^{-25}\leq k \; \xi_{diff}(\eta) \leq 10^{-16}\; ,
\end{equation}
\noindent between reheating and the time of the QCD phase
transition.

The contribution from the hard modes of both the charged scalar
and gauge fields which remain in local thermodynamic equilibrium
lead to an effective mass for the scalar field. This
\emph{thermal} mass is obtained from the long-wavelength limit of
the scalar field self-energy and includes the hard thermal loop
contributions from the gauge and scalar
fields\cite{kapusta,lebellac}. This thermal mass is given by
\begin{equation}\label{thermalmass}
m^2_{T} = \frac{T^2_R}{24} \left(\lambda  + 3 \; e^2\right) \; .
\end{equation}
 Finally, another important quantity is the Debye
screening length that determines the scale at which long-range
forces are screened by the polarizability of the medium. In an
ultrarelativistic plasma, the comoving Debye screening length is
given by\cite{kapusta,lebellac}
\begin{equation}\label{debyelength}
\xi_D \sim \frac{1}{e \; T_R}
\end{equation}
the ratio of the Debye screening length to the Hubble radius is
given by
\begin{equation}
\frac{\xi_D}{d_H} \sim \frac{1}{e} \;  \frac{T(\eta)}{M_*}
\end{equation}
Hence $\xi_D \ll d_H(\eta)$ for $T(\eta) \leq 10^{16}\mbox{Gev}$,
thus long range forces are screened over very short distances. The
formation of long-wavelength domains with typical size of the
order of the Hubble radius\cite{scaling} leading to strong charge
and current fluctuations that will seed magnetic fields, will not
be hindered by long-range forces, which are effectively screened
over sub-horizon distances.

Magnetic field generation via charge asymmetries during a period
in which electromagnetism was spontaneously broken was previously
studied by Dolgov and Silk\cite{dolgovsilk} who argued that
long-range forces would be screened by the Higgs mechanism. This
is different from the situation studied in this article, where the
$U(1)$ symmetry associated with electromagnetism (rather
hypercharge) is \emph{not} spontaneously broken. Long range forces
are screened by the plasma, a situation not considered
in\cite{dolgovsilk}.

\subsection{Scalar fields dynamics}

For completeness and to highlight the  aspects of the
non-equilibrium dynamics most relevant to the generation of
magnetic fields, we summarize the main features of scalar field
dynamics. For further details the reader is referred
to~\cite{nuesfrw,nuestros,destri}. In what follows we will neglect
the backreaction of the gauge fields on the dynamics of the scalar
fields. The rationale for this is that the main non-equilibrium
processes that lead to magnetogenesis will be non-perturbative in
the \emph{scalar} sector and result from the instabilities
associated with the phase transition. The contribution from the
gauge fields, in the form of self-energies for the scalar fields,
do not feature the instabilities associated with the phase
transition and will, furthermore, be suppressed at least by one
power of $\alpha$ the (hyper) electromagnetic coupling constant as
compared to the scalar self-interaction.

As described above, the non-equilibrium evolution of
long-wavelength modes begins with the  spinodal instabilities
which  result in an exponential growth of the amplitudes for
long-wavelength fluctuations. When the non-linearity becomes of
the same order as the tree-level terms in the equations of motion,
the back reaction of these fluctuations  shuts-off the
instabilities~ \cite{nuesfrw,nuestros,destri}.  Therefore a
non-perturbative treatment of the dynamics is required. The large
$N$ limit of the scalar sector allows a systematic
non-perturbative treatment of the dynamics which is renormalizable
and  maintains the conservation laws\cite{nuesfrw,nuestros}.

We will therefore study the dynamics in leading order in the large
$N$ limit that already reveals the important non-equilibrium
features of the evolution.

The contribution from the gauge fields to the equations of motion
of the long-wavelength modes of the scalar fields arise through
self-energy corrections. To lowest order in $\alpha$ these are
dominated by the hard modes of the gauge fields (hard thermal
loops) which lead to a contribution to the thermal mass given by
$eT/\sqrt{8}$~\cite{kapusta,lebellac} and had already been
accounted for in the thermal mass (\ref{thermalmass}).

The \emph{non-equilibrium} effects in the gauge contribution of
the scalar self-energy will arise from polarization loops in the
intermediate photon lines and are, therefore, of
$\mathcal{O}(\alpha^2)$. Since our calculation will be to lowest
order in $\alpha$, we will neglect these contributions. Thus, to
this order the gauge field contribution to the scalar self-energy
is accounted for in the thermal mass.

 Hence the dynamics
of the scalar field is studied along the same lines as presented
in refs.~\cite{destri,nuesfrw,nuestros} but the only difference is
in the initial conditions in the modes that reflect the thermal
mass in LTE.

Since symmetry breaking is chosen along the direction of the
neutral field $\Psi$ we write
\begin{equation} \Psi(\vec x,\eta ) =
\sqrt{N} \; \varphi(\eta ) + \chi(\vec x,\eta ) \quad ; \quad
\langle \chi(\vec x,\eta ) \rangle = 0 \label{expecval}
\end{equation}
\noindent where the expectation value is taken in the time evolved
density matrix or initial state. The leading order in the large N
limit is obtained either by introducing an auxiliary field and
establishing the saddle point or equivalently by the
factorizations\cite{nuesfrw,nuestros}
\begin{eqnarray}
&&(\Phi^{\dagger}\Phi)^2 \rightarrow 2 \langle
\Phi^{\dagger}\Phi\rangle \Phi^{\dagger}\Phi \cr \cr &&\chi
\Phi^{\dagger}\Phi \rightarrow \chi \langle
\Phi^{\dagger}\Phi\rangle \nonumber \; .
\end{eqnarray}
The non-linear terms of the $\sigma$ field lead to contributions
of $\mathcal{O}(1/N)$ in the large $ N $ limit, and to leading
order the dynamics is completely determined by the $ N $ complex
scalars $\Phi_r$. This factorization that leads to the leading
contribution in the large $ N $ limit makes the Lagrangian for the
scalar fields quadratic (in the absence of the gauge coupling) at
the expense of a self-consistent condition: thus charged fields
$\Phi$ acquire a self-consistent time dependent mass.

The dynamics is determined by the Heisenberg equations of motion
of the neutral field $ \Psi$ and the charged fields $ \Phi
$~\cite{nuesfrw,nuestros,destri}.  We will consider that at the
onset of the radiation dominated era, the system is in the
symmetric high temperature phase in local thermal equilibrium with
a vanishing expectation value for the scalar fields. In the
absence of explicit symmetry breaking perturbations the
expectation value of the scalar field will remain zero throughout
the evolution, thus $\varphi\equiv 0$.

It is convenient to introduce the mode expansion of the charged
fields
\begin{equation}
\Phi_r(\eta ,\vec x)= \int \frac{d^3 k}{\sqrt{2 \, (2\pi)^3}}
 \left[ a_r(\vec k) \; f_k(\eta )\;
e^{i\vec k\cdot \vec x}+ b_r^\dagger(\vec k)\; f^*_k(\eta )\;
e^{-i\vec k\cdot \vec x} \right]\quad  , \quad r=1,\ldots,N \; .
\label{phidecompo}
\end{equation}
In leading order in the large $N$ limit, the Heisenberg equations
of motion for the charged fields translate into the following
equations of motion for the mode functions for
$\eta>\eta_R$~\cite{nuesfrw,nuestros,destri}
 \begin{equation}
\left[\frac{d^2}{d\eta
^2}+k^2-M^2(\eta)+\frac{\lambda}{2}\varphi^2(\eta )+
\frac{\lambda}{2N}\langle\Phi^{\dagger}\Phi\rangle\right ] \;
f_k(\eta )=0 \;. \label{unsaledeqnsofmot}
\end{equation}
We must now append initial conditions for  the mode functions
$f_k(\eta )$.  The initial conditions on the mode functions
$f_k(\eta )$ depend on the value of the wavevector $k$ as compared
to the horizon scale $H_R^{-1}$ at the reheating time:
\begin{itemize}
\item{$\mathbf{k>H_R}$:} for fluctuations inside the horizon, we
may assume thermal quasi-particle boundary conditions at the
reheating temperature:
\begin{equation}
f_k(\eta_R)= \frac{1}{\sqrt{W_k}} \quad ; \quad{f'}_k(\eta_R)=
-iW_k~f_k(\eta_R)\; , \quad W_k=\sqrt{k^2+m^2_T}\; ,
\label{iniconds}
\end{equation}
\noindent where the frequencies $W_k$ are quasi-particle
frequencies with thermal mass $m^2_T$ given by
eq.(\ref{thermalmass}) at the reheating temperature,
$$
W_k=\sqrt{k^2+m^2_T} \; .
$$
For these modes the assumption of local thermodynamic equilibrium
is well motivated and we have
\begin{equation}
\langle a_r^\dagger(\vec k) \; a_s(\vec k)\rangle =\langle
b^\dagger_r(k) \; b_s(k)\rangle =\delta_{rs} \; n_k~~;~~
n_k=\frac1{e^{\frac{W_k}{T}}-1}\label{BE.scalars}
\end{equation}
\item{$\mathbf{k<H_R}$:} for superhorizon fluctuations, which are
causally disconnected at the reheating time, we cannot assume a
thermalized distribution. The correct distribution has to  be
derived by following the dynamics from the inflationary stage,
when the fluctuations were well inside the horizon. While a
complete discussion of the initial conditions is left to a
forthcoming article, the case under consideration we will see that
the dependence on the initial conditions is rather weak and only
during the initial stages of the phase transition. For the later
stages, dominated by the scaling solution described below, the
dynamics is \emph{universal} and does not depend on the initial
conditions. We will simply assume that both $W_k$ and $n_k$ have a
finite non-zero limit as $k\to0$ namely the only important
quantities  for the dynamics of long-wavelength fluctuations are
\begin{equation}\label{k->0 bc} \lim_{k\to 0}W_k= W_0\;,\quad
0<W_0<\infty,\quad \lim_{k\to 0}n_k= n_0\;,\quad 0<n_0<\infty\;.
\end{equation}
\end{itemize}

With this choice of the initial state we find the backreaction
term to be given by
\begin{equation}\label{backreaction}
\frac{\lambda}{2N}\langle\Phi^{\dagger} \Phi\rangle =
\frac{\lambda}{4}\int \frac{d^3k}{(2\pi)^3}\;|f_k(\eta
)|^2[1+2n_k]\;.
\end{equation}
This expectation value is ultraviolet divergent, it features
quadratic and logarithmic divergences in terms of an upper
momentum cutoff. The quadratic divergence and part of the
logarithmic divergence (the one proportional to the mass term) are
absorbed in a renormalization of the mass term $\mu^2 \rightarrow
\mu^2_R$ and the remainder logarithmic divergence is absorbed into
a renormalization of the scalar coupling $\lambda \rightarrow
\lambda_R$. While these aspects are not relevant for the
discussion here, they are mentioned for completeness, the reader
is referred to \cite{scaling} for details.

After renormalization the self-consistent field
$\frac{\lambda}{2N}\langle \Phi^\dagger \Phi\rangle$ is subtracted
twice, and is given by (for details see~\cite{scaling} and
references therein)
\begin{eqnarray}\label{JplusI}
\frac{\lambda}{2N}\langle \Phi^\dagger \Phi\rangle &=& \lambda_R
\; \left[J(\eta)+I(\eta)\right] \quad , \quad \lambda_R \; J(\eta)
= \frac{\lambda_R}{4\pi^2}\int_0^{\infty} q^2\;dq \;  {|f_q(\eta
)|^2}~n_q \; , \cr \cr \lambda_R  \;
I(\eta)&=&\frac{\lambda_R}{8\pi^2}\int_0^{\infty} q^2\;dq \;
\Biggr\{{|f_q(\eta)|^2}-\frac{1}{q}
+\frac{\Theta(q-K^2)}{2q^3}\biggl[-\mu^2_R+
\frac{\lambda}{2N}\langle \Phi^\dagger \Phi\rangle\biggr]
        \Biggr\}. \label{gsigma}
\end{eqnarray}
\noindent and the mass and coupling are replaced by their
renormalized counterparts $\mu^2_R;\lambda_R$ respectively. Here
$K$ is an arbitrary renormalization scale. In order to avoid
cluttering of notation we now drop the subscript $R$ for
renormalized quantities, in what follows $\mu;\lambda$ stand for
the renormalized quantities.

The finite temperature term $J(\eta)$ has contributions from short
wavelengths for which the mode functions are of the form
$f_q(\eta) \sim e^{iq\eta}/\sqrt{q}$ and contributions from long
wavelengths. The contribution from short wavelengths is the same
as that in equilibrium in  Minkowski space time and determines the
hard-thermal loop~\cite{kapusta,lebellac} contribution to the
self-energy given by\cite{scaling}
\begin{equation}\label{Jhtl}
J_{HTL}= \frac{T^2_R}{24}
\end{equation}
\noindent where we have used that the short wavelength modes are
in thermal equilibrium at the reheating temperature $T_R$. This
hard thermal loop contribution has been self-consistently
accounted for in the thermal mass of the scalar field
(\ref{thermalmass}).

It is convenient to separate the hard thermal loop component
(\ref{Jhtl}) from eq.(\ref{JplusI}) and define
\begin{equation}\label{sigma}
\lambda\Sigma(\eta)= \frac{\lambda}{8\pi^2}\int_0^{\infty} q^2\;dq
\Biggr\{{|f_q(\eta)|^2}(1+2n_q)-\frac{1}{q}\left[1+
\frac{2}{e^{\frac{q}{T_R}}-1}\right]
+\frac{\Theta(q-K^2)}{2q^3}\biggl[-\mu^2+ \lambda\Sigma(\eta
)\biggr] \Biggr\}.
\end{equation}
After renormalization and in terms of dimensionless quantities,
the non-equilibrium dynamics of the charged scalar fields is
completely determined by the  following equations of
motion~\cite{scaling,nuesfrw,nuestros,destri},
\begin{eqnarray}\label{ecmov}
&&\left[\frac{d^2}{d\eta ^2}+
\mathcal{M}^2(\eta)+q^2+\lambda\Sigma(\eta ) \right]f_q(\eta
)=0~~;~~ f_q(\eta_R)=\frac{1}{\sqrt{W_q}}~~;~~{f'}_q(\eta_R)
=-iW_q~f_q(\eta_R)  \label{modeeqnew}
\end{eqnarray}
\noindent with $W_q$ given by eq. (\ref{mass}) and the effective,
(conformal) time dependent mass is given by
\begin{eqnarray}\mathcal{M}^2(\eta)&=&
C^2(\eta) \; \mu^2\left[\frac{T^2_R}{C^2(\eta) \;
T^2_c}-1\right]\label{mossoft}\\
T^2_c &=& \frac{24 \; \mu^2}{\lambda+3e^2} .\label{crittemp}
\end{eqnarray}
The time dependent mass term $\mathcal{M}^2(\eta)$ includes the
high temperature corrections and clearly displays the cooling
associated with the expansion in the form of a time dependent
effective temperature $T_{eff}(\eta)=T_R/C(\eta)$.  The phase
transition occurs at a time $\eta_c$ when $T_{eff}(\eta_c)=T_c$,
thus for $\eta > \eta_c$ the effective time dependent mass term is
$\mathcal{M}^2(\eta) = M^2(\eta)= -\mu^2 \; H^2_R  \;
\eta^2=-\tilde{\mu}^4  \; \eta^2$ as given by equations
(\ref{mass})-(\ref{mu-tilde}).

The full time evolution of mode functions in a radiation dominated
cosmology has been studied analytically and numerically in detail
in ref.\cite{nuesfrw,scaling}. Here we highlight the most
important features which are necessary ingredients to study
magnetogenesis. The reader is referred to\cite{scaling} for a more
comprehensive discussion.

The are two main dynamical stages in the evolution:

\begin{itemize}
\item{Spinodal stage: this is the stage immediately after the
phase transition which is dominated by spinodal decomposition and
the growth of long-wavelength fluctuations\cite{magfiI}.  This
stage spans the time scale $\eta_c\leq \eta \leq \eta_{nl}$ where
the non-linear time scale $\eta_{nl}$ is determined by (see below)
\begin{equation}\label{etanl} \eta^2_{nl} =
\frac{\lambda\Sigma(\eta_{nl})}{\tilde{\mu}^4 }
\end{equation}
During this stage the back-reaction, determined by the term
$\lambda \; \Sigma(\eta)$, can be neglected and the dynamics is
\emph{linear}. }

\item{Scaling stage: This is a stage in which the non-linearity
encoded by the back-reaction term $\lambda\Sigma(\eta)$ are very
important and compete with the tree level term in the equations of
motion. This stage is described by a scaling solution of the
equations of motion for the modes with small wavevectors and
describes the non-equilibrium relaxation of long-wavelength
fluctuations\cite{scaling,turok,durrer}. }

\end{itemize}

\subsubsection{Spinodal stage}\label{sec:spino}

After the phase transition but before the non-linear time scale
after which the back-reaction becomes important, namely for
$\eta_c \ll\eta<\eta_{nl}$ the time dependent mass term is given
by $\mathcal{M}^2=-\tilde{\mu}^4 \eta^2$, and for weak coupling
$\lambda \ll 1$ we can neglect the back-reaction
$\lambda\Sigma(\eta )$. The equations of motion for the mode
functions during this stage are given by
\begin{equation}\label{parab.cylinder}
\left[\frac{d^2}{d\eta ^2}+q^2- \tilde\mu^4\eta^2\right]f_q(\eta
)=0\quad ;\quad q<\tilde\mu^2\eta
\end{equation}
We note that for $T_c\sim 10^{15} \mbox{GeV}$
\begin{equation}\label{muetac}
\tilde{\mu} \; \eta_c =\left(\sqrt{\frac{\lambda+3\;e^2}{24}}
\frac{M_*}{T_c}\right)^{1/2}\sim 10
\end{equation}
and therefore for $\eta > \eta_c$ we are in the regime
$\tilde{\mu} \; \eta  \gg 1 \; .$ It is clear  that the mode
functions $f_q(\eta)$ will increase exponentially in the band of
\emph{unstable} wavevectors $q<\tilde{\mu}^2\eta$.
Eq.(\ref{parab.cylinder}) can be solved exactly in terms of
Hermite functions\cite{gr}
\begin{equation}\label{hermi}
f_q(\eta )= b_q \; e^{- \frac12( \tilde{\mu} \; \eta)^2} \;
H_{\frac12\left( \frac{q^2}{\tilde{\mu}^2}-1\right)}(\tilde{\mu}
\; \eta) + a_q \; e^{\frac12( \tilde{\mu} \; \eta)^2} \;
H_{-\frac12\left( \frac{q^2}{\tilde{\mu}^2}+1\right)}(i \,
\tilde{\mu} \; \eta)
\end{equation}
where the constants $ a_q $ and $ b_q $ are fixed by the initial
conditions (\ref{ecmov}). For $ \tilde{\mu} \; \eta \gg 1 $ we can
use the asymptotic behavior of the Hermite functions\cite{gr},
$$
H_{\nu}(z) \buildrel{z \gg 1}\over = (2\, z)^{\nu}\left[ 1 + {\cal
O}\left(\frac{1}{z^2}\right) \right]
$$
and we find for the mode functions,
\begin{equation}\label{asimod}
f_q(\eta )\buildrel{\tilde{\mu} \; \eta \gg 1}\over = a_q \;
e^{\frac12( \tilde{\mu} \; \eta)^2} \; \left(\tilde{\mu} \;
\eta\right)^{ -\frac{q^2}{\tilde{\mu}^2}-1} \left[ 1 + {\cal
O}\left(\frac{1}{ \tilde{\mu}^2 \; \eta^2}\right] \right]
\end{equation}
Since the exponentially damped solution becomes negligible the
phases of the mode functions $f_q(\eta )$ {\em freeze}, namely,
they become constant in time and are  slowly varying functions of
$q$ for long wavelengths.

This is very similar to the situation in Minkowski space-time,
where the mode functions however increase as $e^{\mu t}$, i.e.
much slower. In any case the soft ($q\to 0$) modes are the most
amplified at the end of the evolution, therefore, the quantum
fluctuations (\ref{sigma}) are dominated by the lower integration
bound $ q = 0 $.

We notice that the freezing of the long-wavelength mode functions
will play an important role in the discussion about the magnetic
field generation, since it assures the independence of the final
result from the initial particle distribution function, except for
subleading corrections.

The physics of the phase transition is essentially the same as in
Minkowski space-time\cite{magfiI,nuesfrw,nuestros,destri}, since
the exponential growth of modes in the spinodally unstable band
will make the back reaction term $ \lambda \Sigma(\eta) $ begin to
grow and eventually cancel the term $-\tilde{\mu}^4\eta^2$ in the
equations of motion (for $\eta >> \eta_c$ the effective time
dependent temperature vanishes).

This will happen at a \emph{non-linear} time scale defined
by~\cite{nuesfrw,nuestros}
\begin{equation}\label{spin}
\lambda \; \Sigma(\eta_{nl}) = \tilde\mu^4 \; \eta_{nl}^2
\end{equation}
Two important aspects are described by $\eta _{nl}$: i) at this
time scale the phase transition is almost complete since $\lambda
\Sigma(\eta_{nl}) = \tilde\mu^4 \; \eta_{nl}^2$ means that
$\lambda \langle \Phi^\dagger \Phi \rangle /2N =
\tilde\mu^4\eta_{nl}^2$, namely the mean square root fluctuations
in the scalar field probe the manifold of minima of the potential.

 ii) At $\eta  \sim \eta _{nl}$  the mean square root fluctuations
  of the field are of order $M^2(\eta_{nl})
/{\lambda}$ probing the vacuum manifold,  and the non-linearities
become very important. The back reaction $\lambda
\Sigma(\eta_{nl})$ becomes comparable to $M^2(\eta_{nl})$ and the
instabilities shut-off. Thus for $\eta_c< \eta  < \eta _{nl}$ the
dynamics is described by the \emph{linear} spinodal instabilities
while for $\eta>\eta_{nl}$ a full non-linear treatment of the
evolution is required. As it will be discussed below this later
stage is described by the emergence of a scaling solution.

For $\eta_{nl} > \eta \gg \tilde\mu^{-1}$ the asymptotic form
(\ref{asimod}) for the mode functions apply and we find for the
the quantum fluctuations (\ref{sigma}) which dominated by the
lower integration bound $ q = 0 $,
\begin{equation}\label{sigts}
\lambda \; \Sigma(\eta_{nl}) = \lambda \; (1+2\,n_0) \;
\frac{{\tilde\mu}^2 \; |a_0|^2}{32 \; \pi^{\frac52}} \;
\frac{e^{\tilde\mu^2\eta_{nl}^2} }{\tilde\mu \; \eta_{nl} \;
\left[\ln\left({\tilde\mu}\;\eta_{nl}\right)\right]^{\frac32}
}\left[ 1 + {\cal O}\left(\frac{1}{{\tilde\mu}\;\eta_{nl}}
\right)\right] \; .
\end{equation}
\noindent This leads to the following estimate for the spinodal
time for weak coupling~$\lambda$
\begin{equation}\label{spinotime}
\eta^2_{nl} = \frac{1}{\tilde\mu^2}\left[ \ln\left(\frac{32 \;
\pi^{\frac52}}{\lambda \; (1+2\,n_0)\;|a_0|^2\;\tilde\mu}\right) +
\frac32 \; \ln \ln\left(\frac{32 \; \pi^{\frac52}}{\lambda \;
(1+2\,n_0)\;|a_0|^2\;\tilde\mu}\right)  + {\cal O}\left(\ln \ln
\ln\frac{1}{\lambda}\right)\right] \; .
\end{equation}
The important point is that the dependence on boundary conditions
and the initial distribution is solely \textit{logarithmic}, thus
we may expect out predictions to be very robust with respect to
changes of the initial conditions. In particular, the scale factor
at this non-linear time scale is given by
\begin{equation}\label{Cnl} C(\eta_{nl}) = \frac{T_R}{\sqrt{M_* \; T_c}}
\left( \frac{24}{\lambda+3\;e^2} \right)^{\frac14} \;
\sqrt{\ln\frac{1}{\lambda}}\left[1+{\cal
O}\left(\frac{1}{\ln\frac{1}{\lambda}}\right)\right] \; ,
\end{equation}
\noindent where we have used eqs. (\ref{H-T}) and
(\ref{crittemp}).

The amplitude of the long-wavelength modes at the non-linear time,
roughly speaking at the end of the phase transition is
approximately
\begin{equation}\label{endPT}
|f_q(\eta _{nl})| = \sqrt{\frac{32 \; \pi^{\frac52}}{\lambda \;
(1+2\,n_0) \; \tilde\mu }} \;\left[ \ln\frac{32 \;
\pi^{\frac52}}{\lambda \; (1+2\,n_0)}\right]^{\frac14-\frac{q^2}{2
\, {\tilde\mu}^2}}  \; .
\end{equation}
As we will discuss in detail below this non-perturbative scale
will ultimately determine the strength of the magnetic fields
generated during the phase transition.

During the intermediate time regime the equal times correlation
function is approximately \begin{equation}\label{correl} \langle
\Phi^\dagger_{q,a}(\eta )\Phi_{q,b}(\eta )\rangle = \delta_{a,b}
\; |a_q|^2 \; e^{\tilde\mu^2\, \eta^2 } \;
e^{-\left(\frac{q^2}{\tilde\mu^2}+1 \right) \ln[{\tilde\mu}
\eta]}\;.
\end{equation} and its Fourier transform for long wavelenghts is
of the form
\begin{equation}\label{correlation.lenght}
\langle \Phi_a(\vec x,\eta)\Phi_b(\vec 0,\eta)\rangle =
\delta_{a,b} \; |a_0|^2 \; \frac{e^{\tilde\mu^2\,\eta^2
}}{\tilde\mu \,\eta} \;{\tilde\mu}^3 \left[ \frac{\pi}{\ln
{\tilde\mu} \eta}\right]^{\frac32}\; e^{-\frac{\vec
x^2}{\xi^2(\eta )}}
\end{equation}
\noindent which determines the time dependent correlation length
of the scalar field,
\begin{equation}\label{corrlength}
\xi(\eta ) = \frac{2}{\tilde\mu} \;  \sqrt{\ln {\tilde\mu}\;
 \eta } = 2 \, \sqrt{\frac{\ln\left( \sqrt{\mu \; H_R} \, \eta\right)
 }{\mu \;  H_R}} \;.
\end{equation}
This expression is valid in the intermediate time regime
$\eta_c<\eta <\eta_{nl}$ during which the non-equilibrium dynamics
is dominated by the spinodal instabilities. The detailed analysis
of the dynamics in refs.~\cite{nuesfrw,nuestros,destri} and the
discussion of the main features presented above can be summarized
as follows:

\begin{itemize}
\item{At intermediate times $\tilde\mu^{-1}\ll \eta  \leq \eta
_{nl}\sim\tilde\mu^{-1}\sqrt{\ln1/\lambda}$ the mode functions
grow exponentially for modes in the spinodally unstable band
$q<M(\eta)$. The phase of these mode functions \emph{freezes},
namely, becomes independent of time and slowly varying with
momentum.}

\item{ At a time scale determined by the spinodal time the
back-reaction shuts off the instabilities and the phase transition
is almost complete. This can be understood from the following: the
backreaction becomes comparable with the tree-level term (for
$\eta> \eta_R$) when $\frac{\lambda}{2N}\langle \Phi^\dagger \Phi
\rangle \approx \tilde\mu^4\eta^2$. This relation determines that
the mean square root fluctuation of the scalar field probes the
minima of the tree level potential. }

\item{During the spinodal stage the correlation length of the
scalar field grows in time and is given by eq.(\ref{corrlength}).
This is interpreted as the formation of correlated domains that
grow in time, and is the hallmark of the process of phase
separation and ordering. This correlation length will be important
in the analysis of the correlation of magnetic fields later.  }

\item{The large fluctuations associated with the growth of
spinodally unstable modes of the charged fields will lead to
\emph{current} fluctuations which in turn will lead to the
generation magnetic fields. Thus the most important aspect of the
non-equilibrium dynamics of the charged fields during the phase
transition is that large fluctuations of the charged fields
associated with the spinodal instabilities will lead to the
generation of magnetic fields. Since the modes with longer
wavelength are the most unstable the magnetic field generated
through the process of phase separation will be of long
wavelength. Furthermore we expect that the magnetic field
generated by these non-equilibrium processes will be correlated on
length scales of the same order as that of the charged field
above.  }

\end{itemize}

\subsubsection{Scaling stage}\label{sec:scaling}

A remarkable result of the evolution in the asymptotic regime
(when the effective temperature has vanished) found in
ref.~\cite{scaling} is that there is a very precise cancellation
between the tree level term $-\mu^2 C^2(\eta)$ and the back
reaction $\lambda \Sigma(\eta)$ in the equations of motion
(\ref{modeeqnew}). The self-consistency condition requires that
for a radiation dominated cosmology\cite{scaling}
\begin{equation}\label{conscond}
\lambda\Sigma(\eta)-\mu^2
C^2(\eta)\stackrel{\eta\rightarrow\infty}{=}-\frac{15}{4\eta^2}
\end{equation}
In this asymptotic regime the solutions of the equations of motion
\begin{equation}\label{asyeqn}
\left[\frac{d^2}{d\eta^2}+k^2-\frac{15}{4\eta^2}\right]f_k(\eta)=0
\end{equation}
\noindent are given by
\begin{equation}
f_k(\eta) =  \sqrt{\eta} \;\left[ \; A_k
\;\frac{J_2(k\eta)}{k^2}+B_k \; k^2 \;  N_2(k\eta) \;\right]\; .
\end{equation}
This solution can be written in terms of the scaling variable
\begin{equation}\label{scalevar}
x=k\; \eta
\end{equation}
in a more illuminating form
\begin{equation}
f_k(\eta)=A_k\; \eta^{5/2}\; \frac{J_2(x)}{x^2}+B_k\; \frac{x^2
N_2(x)}{\eta^{3/2}}  \; ,
\end{equation}
As discussed in detail in ref.\cite{scaling}, the relevant
integrals are dominated by $x \sim 1$, namely by modes with
wavelength of the order of the Hubble radius, thus  the second
contribution proportional to $N_2(x)$ can be safely neglected at
long times.

For $x \lesssim 1$ in the long time regime we can further
approximate $A_k \sim A_0$ and the asymptotic solution during this
stage is of the \emph{scaling form}
\begin{equation}\label{scaleform}
f_k(\eta)=A_0\; \eta^{5/2}\; \frac{J_2(x)}{x^2}.
\end{equation}
Since for $x \lesssim 3 $ and large time the modes with small
wavevector have the largest amplitudes, these dominate the
backreaction. The very precise cancellation (\ref{conscond}) leads
to the following sum rule~\cite{scaling}
\begin{equation}\label{sumrule}
\frac{\lambda}{8\pi^2}\int k^2 \; dk \left|f_k(\eta)\right|^2 \;
(1+2 \; n_0) \stackrel{\eta\rightarrow \infty}{=} \tilde{\mu}^4 \;
\eta^2
\end{equation}
Since for large $\eta$ the integral is dominated by soft modes
$k\sim\frac{1}{\eta} \rightarrow 0$ the distribution function can
be approximated by $n_0$ and the amplitude by $|A_0|^2$. The sum
rule eq.(\ref{sumrule}) then  leads to the identity
 \begin{equation}\label{A0^2}
|A_0|^2 \; (1+2n_0) =\frac{30\pi^3}{\lambda} \; \mu^2\; H^2_R \; ,
\end{equation}
where we used the integral\cite{gr}
$$
\int_0^\infty \frac{dx}{x^2} \; J_2^2(x)=\frac4{15\pi}\;.
$$
\noindent which is dominated by $x \lesssim 3$.   This is a
remarkable result: the product $|A_0|^2 \; [1+2 \; n_0]$ in the
scaling regime \emph{does not depend on the initial conditions on
the evolution}, namely it is universal in the sense that it is
independent of the previous history through the phase transition.
This is an important result which will play an important role in
the power spectrum of the magnetic fields.

An important consequence of this scaling solution is that the
equal time  two-point correlation function of the scalar field is
given by
\begin{equation}\label{correlation}
\langle \Phi_a(\vec x,\eta)\Phi_b(\vec 0,\eta)\rangle =
\delta_{a,b}~ D(z)~~;~~ z=\frac{|\vec x|}{2\eta}
\end{equation}
\noindent which reveals that the correlation length is given by
the size of the causal horizon\cite{scaling}. The dynamical
evolution during the scaling stage is precisely determined by the
growth of horizon-sized domains\cite{scaling}.

We summarize below the important features of the solutions in the
scaling regime that will be used in the computation of the power
spectrum of the magnetic fields.

\begin{itemize}
\item{For $\eta >> \eta_{nl}$ a scaling regime emerges in which
the mode functions are given by eq. (\ref{scaleform}) with $x=k \;
\eta$. This scaling solution describes  the relaxation of
long-wavelength fluctuations of the charged fields. Again the
phase of these modes \emph{freezes} namely is independent of time.
This is important because this fact will entail that the retarded
self-energy of the transverse photon polarization tensor will give
a subleading contribution to the generation of magnetic fields. }

\item{The sum rule eq.(\ref{sumrule}) constrains the product of
the amplitude times the occupation of the long wavelength scaling
modes to be given by eq.(\ref{A0^2}). }

\item{The scaling solution described above is akin to that found
in \emph{classical} models of formation of topological
defects\cite{turok,durrer}. The scaling regime describes the
evolution of long-wavelength fluctuations and the adjustment of
the spatial correlation length of the scalar field to the Hubble
radius\cite{scaling}. }
\end{itemize}

\subsection{Gauge field dynamics}

In a high temperature plasma a very important aspect that must be
taken into account in the  dynamics of gauge fields is the
electric conductivity, which leads to dissipative processes. As
discussed in \cite{magfiI}, the electric conductivity severely
hinders magnetogenesis, and also introduces the diffusion length
scale which could limit the correlation of the magnetic fields
that are generated.

In Minkowski space in equilibrium the conductivity is obtained
from the imaginary part of the photon polarization and it is
dominated by charged particles of momenta $p\sim T$ in the loop
with exchange of photons of momenta $eT < k \ll
T$~\cite{baym,yaffe}. A careful analysis including Debye
(electric) and dynamical (magnetic) screening via Landau damping
leads to the conclusion that the conductivity is given
by~\cite{baym,yaffe}
\begin{equation}\label{sigmacond}
\sigma = \frac{\mathcal{C} \; N \;  T}{\alpha\ln\frac{1}{\alpha \;
N}}
\end{equation}
\noindent with $N$ the number of charged fields and
$\mathcal{C}\sim \mathcal{O}(1)$.

In an expanding cosmology an in particular during phase
transitions, a more precise assessment of the contributions and
meaning of the conductivity must be provided. As it was discussed
in ref.\cite{magfiI}, the fluctuations of the charged fields
during a phase transition will have very different behavior if the
typical wavevector of these modes is of the order of or smaller
than the symmetry breaking scale or much larger than this scale.

Short wavelength modes, those with typical wave vectors much
\emph{larger than} the symmetry breaking scale are insensitive to
the phase transition and are always in local thermodynamic
equilibrium (LTE). For short wavelength modes deep inside the
horizon, the mode functions are of the free field type $f_q(\eta)
\sim e^{iq\eta}/\sqrt{q}$.

Long wavelength modes, those with wavectors of the order of or
smaller than the symmetry breaking scale undergo critical slowing
down and fall out of equilibrium during the phase transition.
These modes become spinodally unstable during the early stages of
the transition as summarized above and analyzed in detail in refs.
\cite{nuesfrw,nuestros,destri,scaling}.

Thus   the contributions from the charged particle fluctuations to
the photon polarization  must be separated into  two very
different regimes: a) the hard momenta $p \gg |M(\eta)| $
correspond to charge fluctuations that are always in local
thermodynamic equilibrium, b) the soft momenta $p \ll |M(\eta)|$
fall out of equilibrium and undergo long-wavelength spinodal
instabilities and enter the scaling regime.

The contribution from hard momenta will lead to a large
\emph{equilibrium} conductivity in the medium, while the
contribution to the polarization from soft momenta will contain
all the non-equilibrium dynamics that lead to the generation of
electromagnetic field fluctuations.

As the instabilities during the phase transition develop, the
fluctuations of the charged fields will generate non-equilibrium
fluctuations in the long-wavelength components of the electric and
magnetic fields with  the ensuing generation of long-wavelength
magnetic fields. However the large conductivity of the medium will
hinder the generation of electromagnetic fluctuations, hence the
conductivity must be fully taken into account to assess the
spectrum of the magnetic and electric fields generated during the
non-equilibrium stage\cite{magfiI}.

We are interested in the generation of long wavelength magnetic
fields, namely $k<<T_R$ but also $k<< \alpha^2  T_R$, since within
the astrophysical application, the wavelength of interest for
magnetic fields are of galactic scale, while $T_R$ corresponds to
a wavelength at the peak of the CMB which today is
$\lambda_{cmb}\sim 1~\mbox{cm}$. Thus the physical situation
corresponds to studying the photon polarization tensor for
long-wavelength photons.

In equilibrium the long-wavelength and low frequency limit
($k,\omega \rightarrow 0$) of the spatial and temporal Fourier
transform of the transverse polarization is given by
\begin{equation}\label{equi}
\Pi_T(k,\omega)= i \; \omega  \; \sigma
\end{equation}
Thus we write for the full transverse polarization for
long-wavelength electromagnetic fields
\begin{equation}\label{pola}
\Pi_T(\eta,\eta',k)= \sigma \frac{d}{d\eta'} \delta(\eta-\eta')
+\Pi_{noneq}(\eta,\eta',k)
\end{equation}
The first term above includes the contribution from the hard
momentum modes $p \sim T_R$ in the transverse polarization, while
$\Pi_{noneq}(\eta,\eta',k)$ is the contribution from the long
wavelength modes which are unstable in the spinodal stage and take
the scaling form in the scaling regime. Thus in a very well
defined sense, the polarization (\ref{pola}) describes the
effective low energy theory for the transverse photon field.

Our strategy is to obtain the non-equilibrium contribution to the
spectrum of electromagnetic fields to lowest order in $\alpha$ but
treating the conductivity \emph{exactly}.

In a cosmological space-time, the temperature scales with the
inverse of the conformal factor (\ref{Tef})
and therefore the conductivity $\sigma=\sigma(\eta)$ becomes
time-dependent. If we are interested in time scales where the
number of ultrarelativistic charge carriers does not change
significantly, which is the case that we will consider in what
follows, then the time evolution of the conductivity is purely
kinematic (\ref{physisigma}).

An important effect of the conductivity, as discussed in
\cite{magfiI}, is the introduction of a diffusion scale in the
transverse photon propagator. The long-time behavior of the
propagators for the transverse gauge fields: retarded (R),
advanced (A), symmetric (H) in absence of non-equilibrium
contributions and for ultrasoft momenta $k\ll \alpha^2 T$
$$
\mathcal{D}^{(\sigma)ij}_{R,A,H}(\eta,\eta',k)={\cal P}^{ij}({\hat{\bf
p}}) \; \mathcal{D}^{(\sigma)}_{R,A,H}(\eta,\eta',k)
$$
obey\cite{turnerwidrow} (see\cite{magfiI} for details)
\begin{eqnarray}
&&\left[\frac{d^2}{d\eta^2}+k^2+\sigma(\eta) \; C(\eta) \;
\frac{d}{d\eta}
\right]\mathcal{D}^{(\sigma)}_R(\eta,\eta',k)=\delta(\eta-\eta')~~;
~~\mathcal{D}^{(\sigma)}_R(\eta,\eta')=0~~ \mathrm{for}~\eta<\eta'\cr \cr
&&\left[\frac{d^2}{d\eta^2}+k^2+\sigma (\eta) \; C(\eta) \;
\frac{d}{d\eta}
\right]\mathcal{D}^{(\sigma)}_A(\eta,\eta',k)=\delta(\eta-\eta')~~;
~~\mathcal{D}^{(\sigma)}_A(\eta,\eta')=0~~ \mathrm{for}~\eta>\eta'\cr \cr
&&\left[\frac{d^2}{d\eta^2}+k^2+\sigma(\eta) \; C(\eta) \;
\frac{d}{d\eta} \right]\mathcal{D}^{(\sigma)}_{H}(\eta,\eta',k)=0\; ,
\label{con}
\end{eqnarray}
\noindent with the transverse projector
\begin{equation}\label{projector}
{\cal P}_{ij}({\hat{\bf p}})=\delta^{ij}-{\hat{\bf p}}^i{\hat{\bf
p}}^j  \; .
\end{equation}
Due to eq. (\ref{physisigma}) the comoving conductivity
$\sigma_R=\sigma(\eta) \; C(\eta)$ is an \emph{invariant} quantity
in the regime in which the number of ultrarelativistic charge
carriers is constant. The estimate given by eq.(\ref{sigrat})
clearly indicates that during the radiation dominated era between
reheating and the QCD phase transition, $\sigma_R \; \eta\gg1$.

Then for $k\ll \sigma_R$ (which is certainly fulfilled since the
relevant wavevectors are $k\ll T \ll T/\alpha \sim \sigma_R$) and
$\eta\gg 1/\sigma_R$  we can safely neglect the second order time
derivatives in eqs.(\ref{con}), leading to the following
equations,
\begin{eqnarray}
&&\mathcal{D}^{(\sigma)}_R(\eta,\eta',k)=
\mathcal{D}^{(\sigma)}_C(\eta,\eta';k)\;\theta(\eta-\eta') \quad , \quad
\mathcal{D}^{(\sigma)}_A(\eta,\eta',k)=\mathcal{D}^{(\sigma)}_C(\eta,\eta';k)\;
\theta(\eta'-\eta)\label{advasig} \\
&&\mathcal{D}^{(\sigma)}_H(\eta,\eta',k)=
i\;\frac{e^{-\frac{k^2}{\sigma_R}(\eta+\eta')}}{\sigma_R}\
\label{homosig}\quad , \quad \mathcal{D}^{(\sigma)}_C(\eta,\eta';k)=
\frac{e^{-\frac{k^2}{\sigma_R}(\eta-\eta')}}{\sigma_R}\label{DC}
\; .
\end{eqnarray}

\section{Magnetic field spectrum }
As discussed in detail in reference\cite{magfiI}, the quantity of
astrophysical relevance is the correlation function
\begin{equation}\label{corrfuncB} <\hat B^i(\eta ,\vec x) \hat B^i(\eta ,\vec
0)>_{\rho} \; ,
\end{equation}
where the sum on repeated indices is understood. $B(\eta ,\vec x)$
above is a \emph{Heisenberg operator} and the expectation value is
in the initial density matrix. From this quantity, the spectrum of
the magnetic field is obtained in the coincidence limit
\begin{equation}\label{def.spectrum.B} S_B(\eta ,k)=\frac12\lim_{\eta'\to \eta }\int
d^3x <\{\hat B^i(\eta ,\vec x), \hat B^i(\eta ',\vec 0)\}>_\rho
e^{i\vec k\cdot \vec x}\;, \end{equation} where $\{\;,\;\}$
denotes the anti commutator. And from $S_B(\eta ,k)$ we can
extract the \emph{physical} magnetic energy density  stored on
\emph{comoving} length scales larger than a given $L$
\begin{equation}
\Delta\rho_B(L,\eta )= \frac{1}{2\pi^2 }\int_0^{\frac{2\pi}{L}}
k^2 \; S_B(\eta ,k) \; dk   \; .
\end{equation}
where we have restored the powers of the scale factor arising from
the transformation to conformal time. Denoting by $\Delta
\rho_B(L,\eta)$ the contribution from the non-equilibrium
generation (subtracting the local thermodynamic equilibrium
contribution), a quantity of cosmological relevance to assess the
relative strength of the generated magnetic field is given by the
ratio of the power on scales larger than $L$ to the energy density
in the radiation background
\begin{equation}\label{ratio}
r(L,\eta)= \frac{\Delta\rho_B(L,\eta)}{\rho_{\gamma}(\eta)} \; ,
 \end{equation}
\noindent where \begin{equation}\label{CRB} \rho_\gamma
=\frac{\pi^2 T^4_R}{15 }
\end{equation}
is the \emph{comoving} energy density in the thermal equilibrium
background of photons.

The \emph{physical}  energy densities $ \Delta
\rho_{B,phys}(L,\eta) \; , \rho_{\gamma,phys} $ are obtained from
the comoving expressions above by rescaling $\rho \rightarrow
\rho/C^4(\eta)$ as can be seen from the conformal rescaling
(\ref{physicalfields}). Thus the ratio $r(L,\eta)$ would be a
constant in the absence of non-equilibrium generation or
dissipative processes. Hence  the time dependence of the ratio
(\ref{ratio}) only is solely a consequence of  the non-equilibrium
generation mechanisms or dissipative processes (such as magnetic
diffusion in a conducting plasma) but not through the cosmological
expansion.

\subsection{Exact formulation}

In terms of the transverse component of the gauge field
$A^i_T(\eta,\vec k) = \mathcal{P}^{ij}(\vec k) \; A^j(\eta,\vec
k)$ with the transverse projection tensor $\mathcal{P}^{ij}(\vec
k)=\delta_{ij}-\frac{k_i k_j}{\vec k{}^2}$ the spectrum
$S_B(\eta,k)$ can be written as (with an implicit sum over
indices)
\begin{equation} S_B(\eta,k)= -\frac{i}{2} \int d^3x\; e^{i\vec
k\cdot \vec x} \; k^2 \; \left.\mathcal{D}^{ii}_H(\eta,\eta';\vec
x)\right|_{\eta=\eta'} \label{bspectrum} \end{equation} \noindent
in terms of  the \emph{symmetric} correlator of the transverse
gauge field \begin{eqnarray} \mathcal{D}^{ij}_H(\eta,\eta';\vec x)
& = & \mathcal{D}^{ij~>}(\eta,\eta';\vec
x)+\mathcal{D}^{ij~<}(\eta,\eta';\vec x)
\nonumber \\
\mathcal{D}^{ij~>}(\eta,\eta';\vec x) & = & i< A^i_T(\eta,\vec x)
A^j_T(\eta',\vec 0)> ~~; ~~\mathcal{D}^{ij~<}(\eta,\eta';\vec x) =
i< A^j_T(\eta',\vec 0) A^i_T(\eta,\vec x)> \label{symmetcorrA}
\end{eqnarray} Where the correlation functions in
(\ref{symmetcorrA}) are Wightmann functions without time ordering
computed in the initial density matrix. The main reason for
introducing the definitions above is that there is a well
established framework for obtaining these correlation functions in
non-equilibrium quantum field theory as discussed below.

>From $S_B(\eta,k)$ we can extract the magnetic energy density
\begin{equation} \label{magn.energy} \rho_B(\eta)= \int
\frac{d^3k}{(2\pi)^3}\;S_B(\eta,k) \; , \end{equation} where the
ultraviolet behavior is understood to be regulated in some gauge
invariant manner, for example with dimensional regularization.

>From the magnetic energy density we can define an
\textit{effective magnetic field} $B_{eff}(\eta)$ such that
\begin{equation}\label{B.eff} \frac12B^2_{eff}(\eta)\equiv \rho_B(\eta)
\end{equation} In the non-interacting, thermal equilibrium case,
which will be relevant to compare the energy density in the
generated magnetic field to that in the thermal radiation
background, \begin{equation}\label{thermaleq} S_B^{(0)}(k)=k \;
[1+2n(k)]
\end{equation} \noindent  where $n(k)$ is the Bose-Einstein
distribution function. Subtracting the irrelevant vacuum
contribution, the effective magnetic field is then given by the
Stefan-Boltzmann form \begin{equation}\label{SB} \vec
B_{eff}^2{}^{(0)}=\frac{\pi^2T^4}{15}\;. \end{equation} The phase
transition generates a dynamic effective magnetic field $\Delta
B_{eff}(\eta)$ through the interaction between the charged fields
and the electromagnetic field. Hence \emph{a priori} we would
expect that $\Delta B_{eff}(\eta)$ can be obtained systematically
in a power series expansion in the electromagnetic coupling
$\alpha$, namely $\Delta B_{eff}(\eta)= \alpha \Delta
B_{eff}^{(1)}(\eta)+ \alpha^2 \Delta B_{eff}^{(2)}(\eta)+\cdots $.

This would be the case were it not for the fact that  the DC
conductivity is important to estimate reliably the amplitude and
the correlations of the generated magnetic field.  As it will be
discussed below in a high temperature plasma the conductivity is
determined by fluctuations of the  charged fields with typical
momenta $p \sim T$. These short wavelength modes  are in LTE while
only low momentum modes of the charged fields will fall out of
equilibrium and undergo spinodal instabilities during the phase
transition. The DC conductivity is \emph{nonperturbative and
non-analytic} in $\alpha$ and to leading logarithmic order it is
approximately given by\cite{baym,yaffe}
\begin{equation}\label{conductivity2} \sigma = \frac{\mathcal{C}{N}
T}{\alpha\ln\frac{1}{\alpha{N}}}
\end{equation} \noindent with ${N}$ the number of charged fields
and $\mathcal{C}$ a numerical constant of order
one\cite{baym,yaffe}.

Thus while we cannot provide an expression for  the generated
magnetic field as a power series expansion in $\alpha$ because the
presence of the conductivity prevents such expansion, the strategy
that we pursue here  is to treat the long-wavelength
non-equilibrium current fluctuations in perturbation theory in
$\alpha$ while the short wavelength contribution will be accounted
for in the conductivity. This will be discussed below in detail,
but the important point of this discussion is that the
long-wavelength fluctuations that lead to the generation of the
magnetic field will be treated to lowest order in $\alpha$ and to
leading order in the large $N$ limit.

The reliability of this expansion will be guaranteed if
\begin{equation} \frac{\Delta B_{eff}(\eta)}{B_{eff}^{(0)}} \ll 1
\end{equation} In cosmology the important quantity is the energy
density in long-wavelength magnetic fields on scales $ L $ equal
to or larger than the galactic scales. Assuming rotational
invariance, we introduce the \emph{energy} density of the magnetic
field generated by the non-equilibrium fluctuations
\begin{equation}\label{rholongwave} \Delta\rho_B(L) = \frac{1}{2\pi^2}
\int^{2\pi/L}_0 k^2~\Delta S_B(k,t) \; dk\;. \end{equation} The
quantity of cosmological relevance is
\begin{equation}\label{ratio2} r(L)=
\frac{\Delta\rho_B(L)}{\rho_\gamma} \end{equation} \noindent where
$\rho_\gamma =\pi^2 T^4/15$ is the energy density in the thermal
equilibrium background of photons.

As discussed in section IV, the physical magnetic field in a
cosmological background is diluted by the expansion as $B \sim
1/C^2(\eta)$ with $C(\eta)$ the scale factor. Therefore, in the
absence of  processes that generate or dissipate magnetic fields
 the ratio (\ref{ratio2}) would be constant
under cosmological expansion because of the redshift of the
temperature $T \sim 1/C(\eta)$. In a cosmological background the
ratio (\ref{ratio2}) will only depend on time through the
generation or dissipative mechanisms (such as magnetic diffusion
in a conducting plasma) but not through the cosmological
expansion.

For $r \geq 10^{-34}$ the linear (kinetic) dynamo may be
sufficient to amplify cosmological seed magnetic fields and for
$r\geq 10^{-8}$ the collapse of protogalaxies with constant
magnetic flux may be sufficient to amplify the seed magnetic
fields\cite{turnerwidrow,grasso,dolgov2,giova1,Giovanni}.

In the cosmological setting this ratio is approximately constant
if the conductivity in the plasma is very large after the
non-equilibrium dynamics has taken place. The approximations
invoked to estimate the spectrum of the generated magnetic field
will be reliable provided $r(L) \ll 1$.

\subsubsection{Non-equilibrium dynamics of electromagnetic
fluctuations}\label{non.eq.methods}

The formulation of quantum field theory out of equilibrium is by
now well established in the literature and we refer the reader to
refs~\cite{ctp,eri95,eri97} for details.  The generating
functional of real-time non-equilibrium correlation functions can
be written as a path integral along a contour in (complex) time.
The forward and backward branches of this contour represent the
time evolution forward and backward as befits the time evolution
of an initial density matrix\cite{ctp,nuestros}. Gathering all
bosonic (scalar and vector) fields generically in a multiplet
$\mathbf{\Psi}$ the generating functional of correlation functions
is given by~\cite{ctp,eri95,eri97}
\begin{equation}\label{genefunc}
Z(J_{+},J_{-})=\int[\mathcal{D}\Psi_{+}\mathcal{D}\Psi_{-}]\; e^{i
\int d^4x \left\{\mathcal{L}(\Psi_{+})-\mathcal{L}(\Psi_{-})+
J_{+}\Psi_{+}- J_{-}\Psi_{-}\right\}}\;. \end{equation} \noindent
where functional derivatives with respect to the sources $J_{\pm}$
lead to the non-equilibrium real time correlation functions. The
doubling of fields with labels $\pm$ is a consequence of the fact
that the non-equilibrium generating functional corresponds to
forward $(+)$ and backward $(-)$ time evolution and suggests
introducing a compact notation in terms of the  doublet
$(\Psi_{+},\Psi_{-})$ and the  metric in internal space~\cite{ctp}
\begin{equation}\label{cab} c^{ab}=\begin{pmatrix}1&0\cr
0&-1\end{pmatrix}=c_{ab}^{-1}\;, \end{equation} \noindent where
the labels $a,b= \pm$. This notation is particularly useful to
obtain the non-equilibrium version of the Schwinger-Dyson
equations for the propagators~\cite{ctp}.

In the case under consideration the main ingredients for the
non-equilibrium description are the following

\begin{figure}[ht!]
\includegraphics[width=5in,keepaspectratio=true]{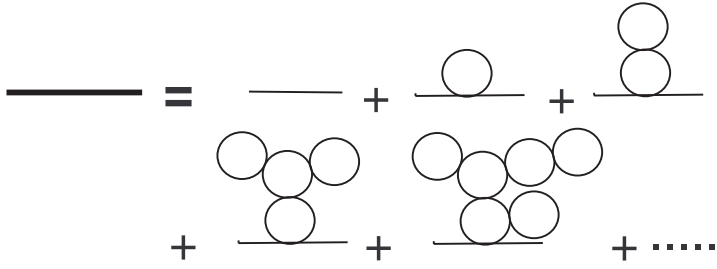}
\caption{ Scalar propagators in leading order in the large N. For
vanishing expectation value of $\sigma$ these are given by a sum
of cactus-type diagrams.} \label{fig:cactus}
\end{figure}

\begin{itemize}
\item{ {\bf Transverse photon propagators}

The real time Green's functions for transverse photons are given
by \begin{equation}\label{photonprops}
{\langle}{A}^{(a)}_{Ti}(\vec{x},t){A}^{(b)}_{Tj}(\vec{x^\prime},
t^{\prime}){\rangle}=-i\int \frac{d^3k}{(2\pi)^3} \;{\mathcal
D}_{ij}^{ab}(k;t,t^\prime)\;e^{-i\vec{k}\cdot(\vec{x}-\vec{x^\prime})}\;,
\end{equation} where the explicit form of ${\mathcal D}_{ij}^{ab}
(k;t,t^\prime)$ is \begin{eqnarray} &&{\mathcal
D}_{ij}^{++}(k;t,t^{\prime})={\cal P}_{ij}(\vec{k}) \;
\left[{\mathcal D}_{ij}^{>}(k;t,t^{\prime})\Theta(\eta-t^{\prime})
+{\mathcal D}_{ij}^{<}(k;t,t^{\prime})\Theta(\eta^{\prime}-t) \right]\;, \label{phot++}\\
&&{\mathcal D}_{ij}^{--}(k;t,t^\prime)= {\cal P}_{ij}(\vec{k}) \;
\left[{\mathcal D}_{ij}^{>}(k;t,t^{\prime})\Theta(\eta^{\prime}-t)
+{\mathcal D}_{ij}^{<}(k;t,t^{\prime})\Theta(\eta-t^{\prime})
\right]\;,
\label{phot--}\\
&&{\mathcal D}_{ij}^{+-}(k;t,t^\prime)={\cal P}_{ij}(\vec{k}) \;
{\mathcal D}_{ij}^{<}(k;t,t^{\prime})\;; \; {\mathcal
D}_{ij}^{-+}(k;t,t^\prime)={\cal P}_{ij}(\vec{k}) \; {\mathcal
D}_{ij}^{>}(k;t,t^{\prime}) \label{photpm} \end{eqnarray} and
${\cal P}_{ij}(\vec{k})$ is the transverse projection operator
(\ref{projector}) } \item{ {\bf Scalar propagators:}
\begin{equation}\label{scalarprops}
{\langle}{\Phi_r}^{(a)\dagger}(\vec{x},t){\Phi_s}^{(b)}(\vec{x^\prime},
t^{\prime}){\rangle}=-i\delta_{rs}\int \frac{d^3k}{(2\pi)^3}\;
G_k^{ab} (\eta,t^\prime) \;
e^{-i\vec{k}\cdot(\vec{x}-\vec{x^\prime})}\;, \end{equation}
\begin{eqnarray}
&&G_k^{++}(\eta,t^\prime)=G_k^{>}(\eta,t^{\prime})\Theta(\eta-t^{\prime})
+G_k^{<}(\eta,t^{\prime})\Theta(\eta^{\prime}-t)\; , \label{gplpl}
\\
&&G_k^{--}(\eta,t^\prime)=
G_k^{>}(\eta,t^{\prime})\Theta(\eta^{\prime}-t)+
G_k^{<}(\eta,t^{\prime})\Theta(\eta-t^{\prime})\;, \label{gpll}
\\
&&G_k^{+-}(\eta,t^\prime)=G_k^{<}(\eta,t^{\prime})~~; ~~
G_k^{-+}(\eta,t^\prime)=G_k^{>}(\eta,t^{\prime}),
\label{gplmin}\\
&&G_k^{>}(\eta,t^{\prime})=\frac i2\left\{[1+n_k]
f_k(\eta)f^*_k(\eta^\prime)+n_k
f_k(\eta^\prime)f^*_k(\eta)\right\}\; ,\label{greater}
\\
&&G_k^{<}(\eta,t^{\prime})=\frac i2\left\{\left[1+n_k
\right]f_k(\eta^\prime)f^*_k(\eta)+n_k
f_k(\eta)f^*_k(\eta^\prime)\right\}\; .\label{lesser}
\end{eqnarray} For the scalar propagators we have used the
expansion in terms of mode functions given by eq.
(\ref{phidecompo}) and the expectation values given by eq.
(\ref{BE.scalars}).  }
\end{itemize}
The scalar propagators given above imply a non-perturbative sum of
cactus-type diagrams when the expectation value of the $\sigma$
field vanishes. These propagators are depicted in  fig.
\ref{fig:cactus}.

\subsubsection{Non-equilibrium Schwinger-Dyson equations}

In this section we derive an {\bf exact} expression for the
spectrum of the magnetic field from the non-equilibrium set of
Schwinger-Dyson equations.

The full non-equilibrium propagator for the photon field is
obtained from the non-equilibrium effective action resulting from
integrating out the charged fields, and which up to quadratic
order in the photon field is given by
\begin{equation}\label{effectiveaction} \Gamma[A^{\pm}]= \int d^4x
d^4y \left\{A^{a}_{i,T}(x)\left[\partial_{\mu}\partial^\mu
\delta^4(x-y) c^{ab}\delta_{ij}+\Pi^{ab}_{ij,T}(x,y)
\right]A^{b}_{j,T}(y) \right\} \end{equation} \noindent with an
implicit sum over all repeated indices.

\begin{figure}[ht!]
\includegraphics[width=2in,keepaspectratio=true]{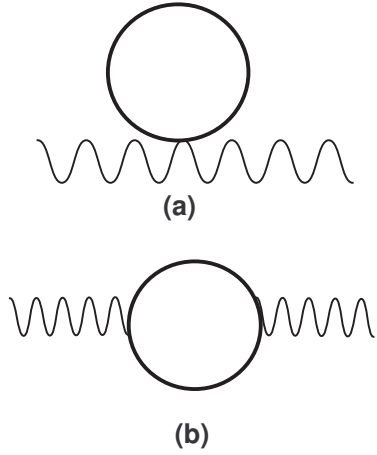}
\caption{ Photon polarization to lowest order in $\alpha$ and to
leading order in the large N. The scalar loops are in terms of the
scalar propagators displayed in fig.\ref{fig:cactus}}
\label{fig:polarization}
\end{figure}

To lowest order in $\alpha$ and to leading order in the large $N$
expansion, the non-equilibrium contribution from the scalar fields
to  the photon polarization is given by the two diagrams shown in
fig. \ref{fig:polarization}.

The tadpole diagram (a) in fig.\ref{fig:polarization} gives a
contribution that is \emph{local} in time, thus we write the
polarization \begin{equation}\label{polar}
\Pi^{ab}_{ij,T}(x,y)={\Pi}^{tad}(\eta) \; c^{ab} \; \delta_{ij} \;
\delta^4(x-y)+ {\mathbf{\Pi}}^{ab}_{ij,T}(\vec x-\vec y,x^0,y^0)
\; , \end{equation} \noindent where the form of the local
(tadpole) contribution ${\Pi}^{tad}(\eta)$ in terms of the metric
$c^{ab}$ is necessary for consistency with the form of the
effective action since the kinetic term is proportional to the
metric~\cite{ctp}. The contribution ${\mathbf{\Pi}}$ is non-local, and in
equilibrium it features absorptive parts. The different components
$a,b = \pm$ are defined in the same manner as the Green's
functions (\ref{phot++})-(\ref{photpm}) and
(\ref{gplpl})-(\ref{gplmin}). The contribution to order $\alpha$
to the non-local part of the polarization is depicted in diagram
(b) in fig.\ref{fig:polarization}.

The photon propagator given in eq. (\ref{photonprops}) is the
inverse of the operator in the quadratic effective action above
and obeys the non-equilibrium version of the Schwinger-Dyson
equation~\cite{ctp}
\begin{equation}\label{SDeqn} \int d^4y
\left[\partial_{\mu}\partial^\mu \delta^4(x-y) \;  c_{ab} \;
\delta_{ij}+\Pi^{ab}_{ij}(x,y) \right]\mathcal{D}^{jk}_{bc}(y,z) =
\delta_i{}^k \; c^a{}_{c} \; \delta^4(x-z)
\end{equation} From the
expressions (\ref{phot++})-(\ref{photpm}) for the different
components of the photon propagator it proves convenient to
introduce the following combinations
\begin{eqnarray}\label{definitions}
&&\mathcal{D}^C_{ij}(x,y)=\mathcal{D}^>_{ij}(x,y)-
\mathcal{D}^<_{ij}(y,x)~~;~~\mathrm{antisymmetric~
propagator} \label{antisym} \\
&&\mathcal{D}^R_{ij}(x,y)=\mathcal{D}^C_{ij}(x,y) \;
\theta(x^0-y^0)~~;~~\mathrm{retarded~
propagator} \label{ret}\\
&& \mathcal{D}^A_{ij}(x,y)=\mathcal{D}^C_{ij}(x,y) \;
\theta(y^0-x^0) ~~;~~\mathrm{advanced~
  propagator} \label{adv}\\
&&
\mathcal{D}^H_{ij}(x,y)=\mathcal{D}^>_{ij}(x,y)+\mathcal{D}^<_{ij}(y,x)~~;~~\mathrm{symmetric~
 propagator}\label{sym}.
\end{eqnarray} \noindent and similar definitions for the polarization
$\mathbf{\Pi}$.

Using spatial translational invariance we can simplify the above
form of the Schwinger-Dyson equations by introducing the spatial
Fourier transform of the transverse propagators
\begin{equation}\label{FT} \mathcal{D}^{ab}_{ij,T}(\eta,\eta';\vec
x-\vec x')=\int\frac{d^3 k}{(2\pi)^3} \; e^{i\vec k\cdot(\vec
x-\vec x')} \; \mathcal{P}_{ij}(\vec k) \;
\mathcal{D}^{ab}(\eta,\eta';\vec k)
\end{equation} and introducing the spatial Fourier transform of the
transverse polarization tensors defined by
\begin{equation}\label{FTSE} {\Pi}^{ab}_{T}(\eta,\eta';\vec x-\vec
x')=\frac{1}{2}\int\frac{d^3 k}{(2\pi)^3} \; e^{i\vec k\cdot(\vec
x-\vec x')} \; \mathcal{P}_{ij}(\vec k) \;
\Pi^{ab}_{ij}(\eta,\eta';\vec k)  \; . \end{equation} We can now
write down the Schwinger-Dyson equations for the transverse photon
propagators.  Since there are only two independent functions
$D^{>,<}$ the Schwinger-Dyson equations (\ref{SDeqn}) can be
simplified to a set of two equations for a particular combinations
of these. Some straightforward algebra leads to the following set
of non-equilibrium Schwinger-Dyson equations (we have suppressed
the subscript $T$ but the equations below refer to the transverse
components) \begin{equation} \label{EQ.DR.k} \int
d\eta_1\left[\left(\frac{d^2}{dt^2_1}+k^2+{\Pi}^{tad}(\eta)
\right)~\delta(\eta-\eta_1)+\Pi_R(\eta,\eta_1;\vec
k)\right]\mathcal{D}_R(\eta_1,\eta';k)=\delta(\eta-\eta')
\end{equation} and
\begin{equation}\label{EQ.DH.k} \int
d\eta_1\left[\left(\frac{d^2}{dt^2_1}+k^2+{\Pi}^{tad}(\eta)
\right)~\delta(\eta-\eta_1)+\Pi_R(\eta,\eta_1;k)\right]
\mathcal{D}_H(\eta_1,\eta';k)=\int
d\eta_1\;\mathbf{\Pi}_H(\eta,\eta_1;\vec
k)\mathcal{D}_A(\eta_1,\eta';k)\;.
\end{equation} \noindent with the definition
\begin{equation}\label{retpol} \Pi_R(\eta,\eta_1;k)= \left[\Pi^>(\eta,\eta_1;\vec
k)-\Pi^<(\eta,\eta_1;k) \right]\Theta(\eta_1-t) \end{equation} A
remarkable aspect of this set of equations is that the right hand
side of eq. (\ref{EQ.DH.k}), namely the inhomogeneity in the set
of equations, only involves the \emph{non-local} contribution to
the polarization, which to lowest order in $\alpha$ is given by
diagram (b) in fig. \ref{fig:polarization}. This is a result of
the form of the local (tadpole) contribution which is proportional
to the metric $c^{ab}$. This is an important point to which we
will come back later in the discussion.

The antisymmetric propagator $ \mathcal{D}_C(\eta,\eta'';k) $ is
the odd solution of the homogeneous evolution equation
\begin{equation} \label{ecDC} \left(\frac{d^2}{d\eta^2}+k^2 +
{\Pi}^{tad}(\eta)\right)\mathcal{D}_C(\eta,\eta'';k)
+\int_{\eta}^{\eta''}dt' \; \Pi_R(\eta,\eta',k) \;
\mathcal{D}_C(\eta',\eta'',k)=0 \; , \end{equation} with the
constraint
\begin{equation} \label{vincu} \left. \frac{
\partial \mathcal{D}_C(\eta,\eta',k)}{\partial \eta} \right|_{\eta' = \eta } = 1
\; . \end{equation} This relation guarantees the correct equal
time canonical commutators.

Since the kernel of the integral equation on the left hand side of
eq. (\ref{EQ.DH.k}) is the same as for the equation that defines
the retarded propagator (\ref{EQ.DR.k}), the solution to eq.
(\ref{EQ.DH.k}) is given by \begin{equation} \label{DH}
\mathcal{D}_H(\eta,\eta';k)=\int d\eta_1\int
d\eta_2\;\mathcal{D}_R(\eta,\eta_1;k) \mathbf{\Pi}_H(\eta_1,\eta_2;\vec
k)\mathcal{D}_A(\eta_2,\eta';k)+F(\eta,\eta';k)
\end{equation} \noindent where $\mathcal{D}_R(\eta,\eta';k)$ is the solution of eq.
(\ref{EQ.DR.k}) and the function $F(\eta,\eta';k)$ symmetric in
$\eta$ and $\eta'$ is a general solution of the homogeneous equation
\begin{equation}\label{H.E.} \int
d\eta_1\left[\left(\frac{d^2}{dt^2_1}+k^2+{\Pi}^{tad}(\eta)
\right)~\delta(\eta-\eta_1)+\Pi_R(\eta,\eta_1;k)\right]F(\eta_1,\eta',k)=0\;.
\end{equation} The homogeneous solution $F$ can be constructed
systematically and its physical significance will be discussed
below.

Now are now in position to provide the final expression for the
 spectrum of the magnetic field. From the expression
 (\ref{bspectrum}) and the solution found above, we find
\begin{equation} S_B(\eta,k)=
-ik^2\left.\mathcal{D}^H(\eta,\eta';k)\right|_{\eta=\eta'}
\label{bspectrumfin} \end{equation} The expression (\ref{DH}) can
be simplified further by taking into account the theta  functions
in the definitions of the retarded and the advanced
($\mathcal{D}_R$ and $\mathcal{D}_A$) propagators [see eqs.
(\ref{ret})-(\ref{adv})] as well as the antisymmetry
$\mathcal{D}_C(\eta,\eta';\vec k)=-\mathcal{D}_C(\eta',t;k)$ of
$\mathcal{D}_C$ leading to
\begin{equation} \label{finalBcor}
S_{B}(\eta,k)=i~k^2\left\{\int_{\eta_R}^\eta d\eta_1\int_{\eta_R}^{\eta}
d\eta_2\;\mathcal{D}_C(\eta,\eta_2;k)\; \mathcal{D}_C(\eta,\eta_1;k) \;
\mathbf{\Pi}_H(\eta_1,\eta_2;k)-F(\eta,t;k)\right\}\;. \end{equation}
Since the product
$\mathcal{D}_C(\eta,\eta_2;k)\mathcal{D}_C(\eta,\eta_1;k)$ is symmetric
in the exchange $\eta_1\leftrightarrow \eta_2$ we can replace
$\mathbf{\Pi}_H(\eta_1,\eta_2;k)$ by $2\mathbf{\Pi}_>(\eta_1,\eta_2;\vec
k)$ and write the final form of the spectrum separating, for
further convenience, the contribution from the inhomogeneous and
homogeneous solutions to $S_B$.
\begin{eqnarray}\label{S.fund}
&& S_B(\eta,k) = S^I_B(\eta,k)+ S^H_B(\eta,k)\nonumber \\
&& S^I_B(\eta, k)=2i~k^2~\int_{\eta_R}^\eta d\eta_1\int_{\eta_R}^{\eta}
d\eta_2\;\mathcal{D}_C(\eta,\eta_2;k)\; \mathcal{D}_C(\eta,\eta_1;k)\;
\mathbf{\Pi}_>(\eta_1,\eta_2;k)\nonumber \\
&&S^H_B(\eta,k)=-ik^2~F(\eta,t;k)\;. \end{eqnarray} \noindent
where $\eta_R$ is  some initial time before the phase transition and
$\mathbf{\Pi}_>$ \emph{does not} include the local tadpole
contributions, it is the \emph{non-local} part of the
polarization.

There is an important aspect associated with the homogeneous
solution $F(\eta,\eta',k)$ and its contribution to the spectrum of
generated magnetic fields $S^B_H(\eta,k)$. This aspect is revealed
by noticing that the \emph{expectation value} of the transverse
gauge field $\mathcal{A}_T(\eta,k)= \langle A_T(\eta,\vec
k)\rangle $, namely the \emph{mean field} obeys the same
homogeneous equation of motion as $F(\eta,\eta',k)$,
\begin{equation}\label{gaugeeqn} \int
d\eta_1\left[\left(\frac{d^2}{dt^2_1}+k^2+{\Pi}^{tad}(\eta)
\right)~\delta(\eta-\eta_1)+\Pi_R(\eta,\eta_1;\vec
k)\right]\mathcal{A}_T(\eta,k)=0 \end{equation} \noindent (we have
suppressed the vector indices to avoid cluttering of notation).

Thus the homogeneous solution $F(\eta,\eta',k)$ can be constructed
out of the independent solutions of the mean field equations of
motion (\ref{gaugeeqn}). The main reason that we bring up this
point is to highlight that the solutions to the mean-field
equations of motion are only \emph{part} of the contributions to
the generation of magnetic fields through non-equilibrium
processes. However, as it will be discussed in detail in the next
section, this contribution can be \emph{neglected} in the present
case of non-equilibrium spinodal decomposition in many
circumstances, and
 the term $S^I_B(\eta,k)$ dominates for late times.

\subsubsection{Electric fields:}

For completeness we now address the generation of electric fields.
The importance of generation of electric fields is mainly related
to the question of equipartition. It is often \emph{assumed} that
the energy density stored in electromagnetic fields is equally
partitioned between electric and magnetic field components, namely
between temporal and spatial gradients. While this is usually the
situation in \emph{equilibrium}, it is not necessarily the case
strongly out of equilibrium such as the situations envisaged in
this article.

The electric field is  the Hamiltonian conjugate field to the
vector potential, and its transverse component is given by
$E_T^i(\eta,\vec k)=-\dot A_T^i(\eta,\vec k)$, therefore the equal
time correlation function of the electric field is given by
\begin{equation} \label{efield} S_{E_T}(\eta,\vec x)=<\dot
A_T^i(\eta,\vec x)\dot A_T^i(\eta,\vec 0>_c=
\partial_\eta\partial_{\eta'}\left.
<\frac12\{A_T^i(\eta,\vec x),A_T^i(\eta',\vec
0)\}>_c\right|_{\eta'=\eta}\;. \end{equation} \noindent where just as
in the case of the magnetic field we wrote the equal time
correlator as the symmetrized connected two-point correlation
function.

Following the steps leading to eq. (\ref{finalBcor}) for the
magnetic field, and using the following identities
$$
\partial_\eta \mathcal{D}_R(\eta,\eta_1;k)=\partial_\eta
\mathcal{D}_C(\eta,\eta_1;k) \; \theta(\eta-\eta_1),\quad
\partial_{\eta'} \mathcal{D}_A(\eta_2,\eta';k)=-
\partial_{\eta'} \mathcal{D}_C(\eta',\eta_2;k) \;
\theta(\eta_2-\eta')
$$
we now find \begin{equation}\label{finalEcor} S_{E_T}(\eta,
k)=\partial_\eta\partial_{\eta'}\left\{2i\int_{\eta_R}^\eta
d\eta_1\int_{\eta_R}^{\eta} d\eta_2\;D_C(\eta,\eta_2;k)D_C(\eta',\eta_1;k)
\Pi_>(\eta_1,\eta_2;k)- F(\eta,\eta',k) \right\}_{\eta=\eta'}
\end{equation} The number of photons produced through the
non-equilibrium processes is given by
\begin{equation}\label{def.N}
N(\eta,k)=\frac{S_{E_T}(\eta,k)+S_B(\eta,k)}{2k}-1\;,
\end{equation} \noindent after summing over the two polarization
states.

The final form for the spectrum of magnetic and electric fields
given by eq. (\ref{S.fund})-(\ref{finalEcor}) as well as the
number of photons given by eq. (\ref{def.N}) are the main tool to
compute the spectrum of electromagnetic fields generated during
non-equilibrium processes and one of the main results of this
article.

We emphasize that these expressions are {\bf exact} and general
and apply in the cosmological setting. They allow to study
the problem of the generation of magnetic fields through
non-equilibrium processes in general situations.

Furthermore, the final expressions for the spectrum of
electromagnetic fluctuations (\ref{S.fund}, \ref{finalEcor}) is
valid more generally in spinor electrodynamics since it only
involves the full polarizations and the Schwinger-Dyson equations
for the correlation functions.

\subsubsection{Spectrum of fluctuations in equilibrium}

Before focusing on the study of the magnetic field generated
during non-equilibrium phase transitions, it is both illuminating
as well as important as a consistency check to address the case of
thermal equilibrium in Minkowski space-time.
In this case the  simplest manner to compute
the spectrum is by using the imaginary time or Matsubara
formulation. where the Euclidean transverse photon propagator is written
as~\cite{kapusta,lebellac}
\begin{equation}
\mathcal{D}^{ij}_E(\tau,\vec x)=<A^i_T(\tau,\vec
x)A^j_T(\tau',\vec 0)>=\int \frac{d^3k}{(2\pi)^3} \;  e^{i{\vec
k}\cdot{\vec x}} \; T \; \sum_{n\in
\mathcal{Z}}\mathcal{P}^{ij}(\vec k) \; \mathcal{D}_T(\omega_n,k)
\; e^{-i\omega_n (\tau-\tau')}~~;~~\omega_n=2\pi T~n
\end{equation} \noindent with
\begin{equation} \mathcal{D}_T(\omega_n,k) = \int dk_0 \;
\frac{\rho_T(k,k_0)}{k_0-i\omega_n}
\end{equation}
\noindent and $\rho_T(k,k_0)$ is the spectral density for
transverse photons which is an odd function of $k_0$. The spectrum
for the magnetic field is obtained from the equal time limit of
the Euclidean propagator, and is therefore given by
\begin{equation}\label{S.ITF}
S_B(k)=2 \; k^2 \;  T\sum_{n\in
\mathcal{Z}}\mathcal{D}_T(\omega_n,k)\;.
\end{equation}
The sum over the Matsubara frequencies can be performed using the
methods described in~\cite{kapusta,lebellac} leading to
\begin{equation}
\label{S.B.eq} S_B(k)=2\int d\omega\; k^2 \; n(\omega) \;
\rho_T(\omega,k)\;. \end{equation} The general form of the
spectral density in terms of the transverse polarization is given
by~\cite{kapusta,lebellac}
\begin{equation} \label{rhoT}
\rho_T(k,k_0) = \frac{1}{\pi}
\frac{\mathrm{Im}\Pi_T(k,k_0)}{\left[k^2_0-k^2-\mathrm{Re}\Pi_T(k,k_0)
\right]^2+\left[\mathrm{Im}\Pi_T(k,k_0) \right]^2}
\end{equation}
The spatial and temporal Fourier transform of the retarded and
advanced transverse photon propagators are given
by~\cite{kapusta,lebellac}
\begin{eqnarray} &&
\mathcal{D}^R_{ij}(k,\omega) =\mathcal{P}_{ij}(k)
\mathcal{D}^R(k,\omega)~~;~~\mathcal{D}^R(k,\omega)=
\frac{1}{-\omega^2+k^2+\mathrm{Re}\Pi_T(k,\omega)+ i \;
\mathrm{Im}\Pi_T(k,\omega)} \label{FTret}\cr \cr &&
\mathcal{D}^A_{ij}(k,\omega) = \mathcal{P}_{ij}(\vec
k)\mathcal{D}^A(k,\omega)~~;~~\mathcal{D}^A(k,\omega)=
\frac{1}{-\omega^2+k^2+\mathrm{Re}\Pi_T(k,\omega)-i \;
\mathrm{Im}\Pi_T(k,\omega)}
\end{eqnarray}
We are now in conditions to establish contact with the
non-equilibrium result of the first section. In equilibrium the
polarization and the propagators are functions of the time
differences and we can then take the Fourier transform in time of
$\mathcal{D}_H(\eta,\eta',k)$ given by eq. (\ref{DH})
\begin{equation}\label{FTDH} \mathcal{D}_H(k,\omega)=
\mathcal{D}_R(k,\omega)\left[\mathbf{\Pi}_>(k,\omega)
+\mathbf{\Pi}_<(k,\omega)\right]\mathcal{D}_A(k,\omega)+
F(k,\omega) \end{equation} \noindent where we have explicitly
written $\mathbf{\Pi}_H$ in terms of $\mathbf{\Pi}_{<,>}$. In
equilibrium the detailed balance (or KMS) condition relates these
components of the polarization to the imaginary
part\cite{kapusta,lebellac}
\begin{equation}\label{KMS}
\mathbf{\Pi}_>(k,\omega)= \frac{i}{\pi} [1+n(\omega)]  \;
\mathrm{Im}\mathbf{\Pi}(k,\omega)~~;~~\mathbf{\Pi}_<(k,\omega)=
\frac{i}{\pi} \;  n(\omega) \;  \mathrm{Im}\mathbf{\Pi}(k,\omega)
\end{equation}
\noindent with $n(\omega)$ being the Bose-Einstein distribution
function.

Furthermore, the Fourier transform of the homogeneous solution
$F(\eta,\eta',k)$ obeys  the Fourier transform of eq. (\ref{H.E.})
namely \begin{equation} \label{FTHOMO}
\mathcal{D}_R^{-1}(\omega,k) \; F(\omega,k)= 0 \end{equation}
Combining all the above ingredients and using the fact that
$\mathrm{Im}\mathbf{\Pi}(k,\omega)$ is an \emph{odd} function of
$\omega$\cite{kapusta,lebellac} we are led to the conclusion that
eq. (\ref{bspectrumfin}) for the spectrum becomes
\begin{equation} \label{SBk}
S_B(k)=\int d\omega\; \left\{2 \; k^2 \; n(\omega) \;
\rho_T(\omega,k) + k^2\; F(\omega,k)\right\}
\end{equation}
\noindent with $\rho_T(\omega,k)$ given by eq. (\ref{rhoT}).

This result differs from  that obtained via the equilibrium
propagator eq. (\ref{S.B.eq}) by  the contribution of the
homogeneous solution $F$.

In order to understand the source of the discrepancy  between the
two formulations, namely the contribution of the homogeneous
solution $F(k,\omega)$, let us focus on the defining equation for
$F(k,\omega)$ (\ref{FTHOMO}).

In order for a non-vanishing solution to this equation to exist,
from the expression for $\mathcal{D}_R(k,\omega)$ we infer that
\begin{eqnarray}
&&\omega^2-k^2-\mathrm{Re}\Pi_T(k,\omega)=0 \label{disper}\\
&&\mathrm{Im}\Pi_T(k,\omega)=0 \label{zerowidth} \end{eqnarray}
Eq. (\ref{disper}) determines the \emph{dispersion relation} of
quasiparticles and eq. (\ref{zerowidth}) determines that these
quasiparticles must have zero width, i.e, the solutions of the
homogeneous equations are \emph{propagating} quasiparticles with
zero width and a dispersion relation given by (\ref{disper}). The
homogeneous solution is therefore \begin{equation} F(k,\omega)
\propto \delta(\omega^2-k^2-\mathrm{Re}\Pi_T(k,\omega))
\end{equation} In a non-perturbative resummation of the Dyson
series for the propagators and the spectral density,  the limit
when $\mathrm{Im}\Pi_T \rightarrow 0^\pm$ leads to
\begin{equation} \rho_T \rightarrow
\mp\delta(\omega^2-k^2-\mathrm{Re}\Pi_T(k,\omega)) \end{equation}
\noindent and we recognize in this case that the possible
contributions of the homogeneous solutions are  accounted for in
the expressions (\ref{rhoT}) and arise from the propagating
states, and to zeroth order account for the contribution to the
spectrum of the magnetic field from \emph{free} photons.

However, in a perturbative expansion perturbation theory is only
reliable \emph{away} from the single particle poles in the
propagator and the limit $\mathrm{Im}\Pi_T \rightarrow 0^\pm$ will
miss the contribution from the isolated poles. The homogeneous
solution $F(k,\omega)$ gives the required contribution, thus
guaranteeing the consistency of the perturbative expansion. Thus
while the homogeneous solution $F$ \emph{must} be included in a
perturbative calculation of the spectrum, it will be accounted for
in a non-perturbative computation that includes a Schwinger-Dyson
resummation for the full propagator and should \emph{not} be
included in the computation of the spectrum.

In a plasma the true degrees of freedom are \emph{quasiparticles},
in particular for the power spectrum of  magnetic and transverse
electric fields, the important degrees of freedom are transverse
\emph{plasmons}. In the hard thermal loop
approximation\cite{kapusta,lebellac} the transverse plasmons
contribute to the spectral density $\rho_T(k,\omega)$ a pole term
of the form $Z_T(k)[\delta(\omega-\omega_p(k))-
\delta(\omega+\omega_p(k)]$ with $\omega_p(k)$ being the plasmon
dispersion relation\cite{kapusta,lebellac} with $\omega_p(0)=
eT/3$. For $k \gg eT, \; \omega_p(k)\sim k~;~Z_T\sim 1/2k$ and we
recover the power spectrum of free field theory. However for $k
\lesssim eT$ the power spectrum of the magnetic field reveals the
presence of collective degrees of freedom, $S_B(k) \sim
kZ_T(k)[1+2 \, n(\omega_p(k)]$ for $e^2T \leq k \leq eT$.

For ultrasoft frequency and momenta $k,\omega \ll \alpha^2T$ which is the
case of relevance for the cosmological case, the effective low
energy form of the gauge field (transverse) propagator (well below
the plasmon pole) is
\begin{equation}
\mathcal{D}^{(\sigma)}_R(k,\omega) \simeq \frac{1}{\omega^2-k^2-i\sigma \omega}
\end{equation}
leading to the small frequency, long-wavelength form of the
spectral density
\begin{equation}
\rho_T(k,\omega) \simeq \frac{1}{\pi} \frac{\sigma \;
\omega}{(\omega^2-k^2)^2+(\sigma \omega)^2} \; .
\end{equation}
Therefore, for $k \ll \sigma $ and $\omega \ll k$
\begin{equation}
k^2 \; n(\omega) \; \rho_T(k,\omega) \simeq \frac{T}{\pi}
\frac{\Gamma}{\omega^2+\Gamma^2} ~~;~~ \Gamma \equiv
\frac{k^2}{\sigma} \; .
\end{equation}
\noindent Thus for long wavelengths $k\ll \sigma \approx T/\alpha$
the spectral density features a pole at zero frequency leading to
a long-wavelength power spectrum
\begin{equation}\label{SBT}
S_B(k\ll \sigma) = T \; ,
\end{equation}
coming from the first term in eq.(\ref{SBk}).

The plasmon pole contributes to $ S_B(k) $ through the homogeneous
solution $F$ in  eq.(\ref{SBk}) yielding,
\begin{equation}\label{SBp}
S_{B plasmon} \simeq \frac{k^2}{\omega_p(0)}[1+2n(\omega_p(0)]\sim
\frac{2 k^2 T}{\omega^2_p(0)} \sim \frac{k^2}{e^2T}
\end{equation}
where we have used the long-wavelength limit of the dispersion
relation $\omega_p(k)$ and residue
$Z_T(k)$\cite{kapusta,lebellac}. The contribution (\ref{SBp}) is
clearly much smaller than (\ref{SBT}) in the long-wavelength limit
$k\ll eT$. Hence, the magnetic field power spectrum in equilibrium
is $S_B(k) \simeq T$.

Having clarified the role of the homogeneous solution and the form
of the power spectrum within the more familiar equilibrium
setting, we return to the non-equilibrium situation.

\subsubsection{Spectrum out of equilibrium}

Using the results obtained above, in particular the final
expression for the magnetic field spectrum, eqn.~\ref{S.fund} we
are now ready to study the spectrum of the magnetic field
generated by the non-equilibrium evolution of the charged scalar
field.

While the final result for the spectrum given by eqn.
(\ref{S.fund}) is \emph{exact},  in order to make progress and
obtain an estimate for magnetic and electric field generation we
have to make approximations. In what follows we will obtain the
spectra for magnetic fields to lowest order in $\alpha$ and to
leading order in large $N$.

To this order the tadpole (local) and bubble (non-local)
contribution to the polarization are given respectively by
diagrams (a) and (b) in fig. \ref{fig:polarization}. Their
explicit expressions are given by \begin{eqnarray}
\Pi^{tad}(t) &=& -ie^2 N \int \frac{d^3q}{(2\pi)^3} \;
{G}_>(t,t,q) \; , \cr \cr \Pi_{ij,>}(\eta_1,\eta_2,\vec k)&=&ie^2
N \int \frac{d^3q}{(2\pi)^3} \; (2q_i+k_i) \; (2q_j+k_j) \;
G_>(\eta_1,\eta_2,q) \; G_>(\eta_1,\eta_2,|\vec q+\vec k|)\;.
\end{eqnarray} \noindent which leads to the transverse components
\begin{eqnarray}\label{PITranv}
 && \Pi_>(\eta_1,\eta_2, k) =2ie^2 N\int
\frac{d^3q}{(2\pi)^3} \; q^2 \; (1-\cos^2
\theta)~G_>(\eta_1,\eta_2,q) \; G_>(\eta_1,\eta_2,|\vec q+\vec
k|)\; , \cr \cr &&  \Pi_<(\eta_1,\eta_2, k)=\Pi_>(\eta_2,\eta_1,
k)
\end{eqnarray} where $ \cos\theta\equiv {\hat q}\cdot {\hat k} $.
The scalar propagator $G_>(\eta,\eta')$ in terms of the mode
functions $f_q(t)$ that satisfy the evolution equations
(\ref{unsaledeqnsofmot}) is given by [see eq. (\ref{greater})],
\begin{equation}\label{scalpropa} G_>(\eta_1,\eta_2;k)=\frac
i2\left[(1+n_k)f_k(\eta_1)f_k^*(\eta_2)+n_k
f_k^*(\eta_1)f_k(\eta_2)\right]\;.
\end{equation} \noindent and we assumed that before the phase transition the
scalar fields have occupation $n_k$. Inserting this expression in
$S^I_B(\eta,k)$ in eqn. (\ref{S.fund}) we find~\cite{magfiI}

\begin{eqnarray}\label{S.B.longexpr}
&&S^I_B(\eta,k)=e^2 N\int \frac{d^3q}{(2\pi)^3}
q^2(1-\cos^2\theta)  \; 
\left[(1+n_q)(1+n_{|\vec q+\vec k|})\left|\int_{\eta_R}^{\eta}
d\eta_1\;k\;\mathcal{D}_C(\eta,\eta_1,k ) f_q(\eta_1)f_{|\vec
q+\vec k|}(\eta_1)\right|^2+\;\right.\nonumber\\
&&\left.+(1+n_q)n_{|\vec q+\vec k|}\left|\int_{\eta_R}^{\eta}
d\eta_1\;k\;\mathcal{D}_C(\eta,\eta_1,k ) f_q(\eta_1)f^*_{|\vec
q+\vec k|}(\eta_1)\right|^2+\; n_q(1+n_{|\vec q+\vec
k|})\left|\int_{\eta_R}^{\eta}
d\eta_1\;k\;\mathcal{D}_C(\eta,\eta_1,k ) f^*_q(\eta_1)f_{|\vec
q+\vec k|}(\eta_1)\right|^2+\;\right.\nonumber\\
&&\left. +n_qn_{|\vec q+\vec k|}\left|\int_{\eta_R}^{\eta}
d\eta_1\;k\;\mathcal{D}_C(\eta,\eta_1,k ) f^*_q(\eta_1)f^*_{|\vec
q+\vec k|}(\eta_1)\right|^2\;\right] \; .
\end{eqnarray}

\begin{figure}[ht!]
\includegraphics[width=4.00in,keepaspectratio=true]{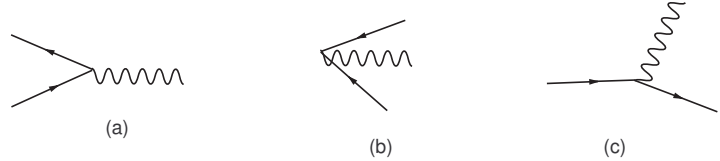}
\caption{ Processes described by $S^I_B$. (a): pair annihilation,
(b): pair production, (c): bremsstrahlung} \label{fig:processes}
\end{figure}

The different terms in this expression have a simple kinetic
interpretation. The first term, proportional to
$(1+n_q)(1+n_{|{\vec q}+{\vec k}|})$ corresponds to the process of
pair \emph{creation} along with the creation of a photon, the
second and third terms, proportional to the combinations $(1+n)n$
correspond to the process of \emph{bremsstrahlung}, and the last
term proportional to $n~n$ corresponds to pair annihilation with
the emission of a photon. These processes are depicted in figure
\ref{fig:processes}.

These processes are forbidden in equilibrium by energy momentum
conservation, however in an expanding cosmology and strongly out
of equilibrium these processes are not only allowed but they give
the leading order contribution to magnetogenesis.

The contribution from the homogeneous solution $S^H_B(\eta,k)$ is
given by

\begin{equation}\label{Shomo} S^H_B(\eta,k) =
-ik^2F(\eta,\eta;k) \; ,
\end{equation}
\noindent where $F(\eta,\eta';k)$ satisfies the homogeneous
differential equation
\begin{equation}\label{Feqn}
\left[\frac{d^2}{d\eta^2}+k^2+\sigma(\eta)C(\eta) \frac{d}{d\eta}
\right]F(\eta,\eta',k)+\int
d\eta_1\;\left[\Pi^{tad}(\eta_1)\delta(\eta-\eta_1)+\Pi_R(\eta,\eta_1)
\right]F(\eta_1,\eta',k)=0 \; ,
\end{equation}
\noindent with $\Pi^{tad}(\eta) , \; \Pi_R(\eta,\eta')$ being the one
loop tadpole (local) and retarded (non-local) contributions
transverse polarization\cite{magfiI}.

We note that the function $F(\eta,\eta';k)$ obeys the same
equation as the transverse gauge mean field\cite{magfiI}, but as
it will be argued in detail below, its contribution to the
spectrum of primordial magnetic fields generated during the phase
transitions is subleading in the scalar coupling constant
$\lambda$. The equation of motion (\ref{Feqn}) can be solved
systematically in an expansion in powers of the non-equilibrium
polarization,
\begin{eqnarray}\label{pertF}
&&F(\eta,\eta;k)  =
F^{(0)}(\eta,\eta;k)+F^{(1)}(\eta,\eta;k)+{\cal
 O}(\alpha^2)  \; , \nonumber \\
 &&F^{(0)}(\eta,\eta',k)=\mathcal{D}_H(\eta,\eta',\vec k) \; ,\nonumber \\
 &&F^{(1)}(\eta,\eta',k)=   \int^\eta_{\eta_R}d\eta_1\;
\mathcal{D}_C(\eta-\eta_1,\vec k) \int^\eta_{\eta_R} d\eta_2
\left[ {\Pi}^{tad}(\eta_2)
~\delta(\eta_1-\eta_2)+\Pi_R(\eta_1,\eta_2;\vec k) \right]
\mathcal{D}_H(\eta_2-\eta',\vec k)+ \nonumber \\& & +\quad
(\eta\leftrightarrow \eta') \; ,
\end{eqnarray}
\noindent where $\Pi^{tad}~ \;~ , \Pi_R$ are the tadpole (local) and
the retarded contribution from the one-loop transverse photon
polarization respectively (for details see\cite{magfiI}).

These are given by ~\cite{magfiI}
\begin{eqnarray}\label{leadtad}
&&\Pi^{tad}(\eta) = -ie^2 N \int \frac{d^3q}{(2\pi)^3} \;
{G}_>(\eta,\eta,q) \quad  ,  \quad \Pi_{R}(\eta_1,\eta_2,\vec
k)=\left[\Pi^>(\eta_1,\eta_2,\vec k)-\Pi^<(\eta_1,\eta_2,\vec
k)\right]\Theta(\eta_1-\eta_2)
\end{eqnarray}
\noindent with
\begin{eqnarray}
 && \Pi_>(\eta_1,\eta_2, k) =2ie^2 N\int
\frac{d^3q}{(2\pi)^3} \; q^2 \; (1-\cos^2
\theta)~G_>(\eta_1,\eta_2,q) \; G_>(\eta_1,\eta_2,|\vec q+\vec
k|)\; , \; 
\Pi_<(\eta_1,\eta_2, k)=\Pi_>(\eta_2,\eta_1, k)\; .\nonumber
\end{eqnarray}
The scalar propagator $G_>(\eta,\eta';k)$ is constructed from the
mode functions $f_q(\eta)$ that satisfy the mode equations
(\ref{parab.cylinder}) and is given by (\ref{scalpropa}); therefore
\begin{equation}\label{piretima}
\Pi_{R}(\eta_1,\eta_2,\vec k)= 4e^2 N\int \frac{d^3q}{(2\pi)^3} \;
q^2 \; (1-\cos^2 \theta)~\mbox{Im}\left[G_>(\eta_1,\eta_2,q) \;
G_>(\eta_1,\eta_2,|\vec q+\vec k|)\right]\Theta(\eta_1-\eta_2) \;
.
\end{equation}
This expression for the retarded self-energy must be contrasted
with that of the contribution from $S^I_B(\eta,k)$ which requires
the \emph{real} part $\mbox{Re}\left[G_>(\eta_1,\eta_2,q) \;
G_>(\eta_1,\eta_2,|\vec q+\vec k|)\right]$. This is an important
difference, the long wavelength modes of largest amplitude in
either phase given by (\ref{asimod}) or by (\ref{scaleform}) are
such that their phases are frozen, namely they do not depend on
time, therefore the products $f_q(\eta_1)f_{|{\vec p}+{\vec
q}|}(\eta_2)$ with only the growing mode solutions are \emph{real}
and  such products will contribute only to $S^I_B(\eta,k)$. This
freezing of phases is a consequence of the classicalization of the
scalar field fluctuations\cite{scaling}.

\bigskip

We now argue that the contribution from $S_B^H$ is subleading.
First of all, the term $F^{(0)}(\eta,\eta;k)$ in eq.(\ref{pertF})
is the solution of the homogeneous equation in absence of
non-equilibrium fluctuations and leads to the local thermodynamic
equilibrium contribution to the power spectrum, which is
independent of the non-equilibrium generation mechanisms. This
contribution has been analyzed in detail in ref.\cite{magfiI} and
will be subtracted. In what follows we focus solely on the
contribution from the non-equilibrium fluctuations.

For intermediate times after the phase transition during the
\emph{spinodal stage} $\eta_c\leq \eta <\eta_{nl}$,  the
long-wavelength mode functions are approximately given by
eq.(\ref{asimod}).

Near the end of the phase transition for $\eta \sim \eta' \sim
\eta_{nl}$ the leading order time dependence of the scalar Green's
functions is approximately given by
\begin{equation}\label{GFts}
f_k(\eta)f_k(\eta') \propto\frac{1}{\lambda} \; .
\end{equation}
where we used eq.(\ref{endPT}). Thus, the contribution from the
tadpole (local term in the self-energy) is of the order
\begin{equation}\label{tadts}
\Pi^{tad}(\eta)_{\eta\sim \eta_{nl}} \propto \frac{e^2}{\lambda} +
\mathrm{subleading} \; .
\end{equation}
This estimate is consistent with the fact that the tadpole
contribution is $ e^2 <\Phi \Phi> $ and near the end of the phase
transition the mean square root fluctuations of the scalar field
probe the vacuum state, namely $<\Phi^2> \sim \mu^2 /\lambda$.
Since the phases of these modes are frozen, there is no
contribution from the leading order to the retarded polarization,
since it requires the imaginary part of the product of propagators
as displayed in eq. (\ref{piretima}). Because of this cancellation
of the leading term, the contribution from the retarded
polarization bubble is of the same order as that of the
tadpole~\cite{Boyanovsky:1999jh,magfiI}.
\begin{equation}\label{pireta}
\Pi_R \propto \frac{e^2}{\lambda} \; .
\end{equation}
A similar argument based on the sum rule (\ref{sumrule}) leads to
the same conclusion in the scaling regime.

The contribution from $S^I_B$ is in both cases of ${\cal
O}(1/\lambda^2)$ since  each long-wavelength mode function is of
order $1/\sqrt{\lambda}$ at the end of the spinodal stage or, by
the sum rule (\ref{sumrule}) in the scaling regime. Thus we can
safely neglect the contribution from $S^H_B$ to the magnetic
spectrum.

 \bigskip

The general expression (\ref{S.B.longexpr}) can be further simplified
by noticing that during most of the phase transition the mode functions
are frozen: therefore the four terms in (\ref{S.B.longexpr}) can be
collected together and the leading contribution to the power spectrum
generated by non-equilibrium fluctuations can be written as
\begin{equation}\label{SButil}
S_B(\eta,k)=(1+2n_0)^2 \;\frac{\alpha N~k^2}{\pi ~\sigma^2_R} \;
e^{-\frac{2k^2}{\sigma_R}\eta} \; \int_0^{\infty} q^4 dq~ d(\cos
\theta) \;
(1-\cos^2\theta)\left|\int_{\eta_R}^{\eta}e^{\frac{k^2}{\sigma_R}\eta_1}\;
f_q(\eta_1)\; f_{|\vec q+\vec k|}(\eta_1)\; d\eta_1\right|^2 \; .
\end{equation}
where $ \theta $ is the angle between the vectors $ \vec q $ and $
\vec k $ and where we have replaced
$$
(1+2n_q)(1+2n_{|\vec q+\vec k|})\simeq (1+2n_0)^2 \; ,
$$
since as highlighted  in section (\ref{sec:spino}) the
 dynamics during both the spinodal stage as well as the scaling stage is
 dominated by the long-wavelength  modes that acquire non-perturbatively large
  amplitudes.

  The final form of the power spectrum generated by the
  non-equilibrium dynamics given by eq.(\ref{SButil}) is the
  basis for the study of primordial magnetogenesis during the
  different stages after the phase transition.

\subsection{Magnetogenesis during the spinodal stage}

The long-wavelength mode functions in the spinodally unstable band
are given by the expression (\ref{asimod}).

The integral over $\eta_1$ for large $\eta$  can be computed
integrating by parts in eq.(\ref{SButil}) as an expansion in
$1/(\tilde{\mu}\eta)^2$. The integral is dominated by the upper
limit, which leads to the cancellation of the exponentials that
contain the conductivity.

The integrals over momenta and angles in eq.(\ref{SButil}) can be
done straightforwardly when the mode functions are given by
eq.(\ref{asimod}). Thus from eq.(\ref{SButil}) whe obtain the
following expression for the spectrum  of magnetic fields
generated by the non-equilibrium fluctuations
\begin{equation}\label{finSBhiT}
S_B(k,\eta\sim \eta_{nl}) = \frac{512 \, \pi^{\frac{9}{2}} \; N
\;\alpha \; k^2 }{\lambda^2\;\sigma^2_R \; {\tilde\mu}^4
\;\xi^5(\eta_{nl})}~ \; e^{-\frac{1}{4}k^2 \xi^2(\eta_{nl})}
\;\left[ 1 + {\cal O}\left(\frac{1}{\ln\frac{1}{\lambda}}\right)
\right]\;.
\end{equation}
where $ \xi(\eta) $ is given by eq.(\ref{corrlength}). In
obtaining this result we used the following
\begin{equation}
\frac{[1+2\,n_0]^2\;|a_0|^4}{\left({\tilde\mu}\eta_{nl}\right)^6}
\; e^{2\left({\tilde\mu}\eta_{nl}\right)^2} = \frac{1024 \;
\pi^5}{\lambda^2 \; {\tilde\mu}^2}
\end{equation}
[see eq.(\ref{spinotime})] and the identities\cite{gr},
\begin{equation}
\int_{-1}^{+1}dx \; (1-x^2)\; e^{-\frac{4 \; q \; k  \;
x}{{\tilde\mu}^2} \; \ln{\tilde\mu}\eta } =
\frac{{\tilde\mu}^6}{16 \; (q \; k \; \ln{\tilde\mu}\eta
)^3}\left\{\frac{4 \; q \; k }{{\tilde\mu}^2} \;
\ln({\tilde\mu}\eta)  \cosh\left[\frac{4 \; q \; k}{{\tilde\mu}^2}
\; \ln{\tilde\mu}\eta  \right] -\sinh \left[\frac{4 \; q \;
k}{{\tilde\mu}^2} \; \ln{\tilde\mu}\eta  \right] \right\}
\end{equation}
and
\begin{equation}
\int_0^{\infty} q \; dq \;  e^{-\xi^2 \; q^2} \left[ \xi^2 \; q \;
k \; \cosh(\xi^2 \; q \; k) - \sinh(\xi^2 \; q \; k) \right] =
\frac{\sqrt\pi}{8}\; k^3 \;\xi \; e^{\frac{1}{4}k^2 \xi^2} \; .
\end{equation}

Notice that the magnetic field  spectrum (\ref{finSBhiT}) is
independent on the amplitude $|a_0|$ and on the initial occupation
$(1+2n_0)^2$. Therefore this result is  quite robust.

This result is  the same as for the Minkowski space-time (see
eq.(7.47) in ref.\cite{magfiI}), except for a multiplicative
factor $ {\tilde\mu}^6 \; \xi^6(\eta_{nl}) \simeq 8 \;
\ln^3\left(\ln\frac{1}{\lambda}\right) $ and  the expression for
the correlation length in the radiation dominated universe
(\ref{corrlength}).

As in Minkowki space-time, the presence of a high conductivity
plasma severely hinders the generation of magnetic fields.
However, a noteworthy aspect is that up to the non-linear time the
magnetic field is still correlated over the size of the scalar
field domains rather than the diffusion length $\xi_{diff} \approx
\sqrt{\eta/\sigma_R}$. The diffusion scale  determines the spatial
size of the region in which magnetic fields are correlated in the
\emph{absence} of non-equilibrium generation. The ratio between
the domain size $\xi(\eta)$ given by (\ref{corrlength}) and the
diffusion length scale $\xi_{diff}(\eta)$ is given by
\begin{equation}\label{scalesratio}
\frac{\xi(\eta_{nl})}{\xi_{diff}(\eta_{nl})} \simeq
\frac{2}{\tilde\mu} \sqrt{\frac{\sigma_R \;
\ln({\tilde\mu}\eta_{nl})}{\eta_{nl}}}\sim
 \left(\frac{M^2_{Pl}\ln\frac{1}{\lambda}}{\mu^2 } \right)^{\frac{1}{4}}\gg 1\; .
\end{equation}
Where we have used the relations (\ref{H-T}), (\ref{Mstar}),
(\ref{comovingsigma}) and (\ref{spinotime}). Thus an important
conclusion of this study is that the magnetic fields generated via
spinodal decomposition are correlated over regions comparable to
the size of scalar field domains which are \emph{much larger} than
the diffusion scale.

The spectrum for the  electric field  can be obtained from that of
the magnetic field by  simply replacing $k\;\mathcal{D}_C
\rightarrow \dot{\mathcal{D}}_C$. In the soft regime and for time
scales $\frac{1}{\sigma_R}\ll \eta\ll \frac{\sigma_R}{k^2}$ we
have $ \dot{\mathcal{D}}_C \simeq -k^2/\sigma_R^2 $ whereas $k
\,\mathcal{D}_c\simeq k/\sigma_R$. Therefore the electric field
spectrum is suppressed by a factor $k^2/\sigma_R^2$ with respect
to the magnetic field, namely
\begin{equation}\label{hiTelec}
S^{\sigma_R}_E(\eta,k)=\frac{k^2}{\sigma_R^2} \;
S^{\sigma_R}_B(\eta,k)\;.
\end{equation}
Thus, in a high temperature plasma with large conductivity the
non-equilibrium processes favor the generation of magnetic photons
instead of electric photons, and again equipartition is not
fulfilled.

The energy density on large scales $\geq L$ again can be computed
in closed form in the limits $L\gg \xi(\eta_{nl})$ and $L\ll
\xi(\eta_{nl})$. We find in the first case from
eq.(\ref{magn.energy}),
\begin{equation}\label{specLcond}
\Delta\rho_B(\eta_{nl},L)=\frac{2^{13} \; \pi^{\frac{15}{2}} }{5
\; \lambda^2} \; \frac{N \; \alpha}{[ {\tilde\mu} \;
\xi(\eta_{nl})]^4 \; \sigma_R^2 \; \xi(\eta_{nl}) \;
L}\;\frac{1}{L^4} ~~;~~ L \gg   \xi(\eta_{nl})\;.
\end{equation}
We find for the opposite case,
\begin{equation}\label{Lchico}
\Delta\rho_B(\eta_{nl},L)=\frac{3 \times 2^{10} \; \pi^3  \; N \;
\alpha}{ \lambda^2 \; {\tilde\mu}^4 \;\sigma_R^2 \;
\xi^{10}(\eta_{nl})} ~~;~~ L \ll   \xi(\eta_{nl})\;.
\end{equation}

The ratio of the magnetic energy density on scales larger than $L$
at the spinodal time and the total radiation energy given by the
Stefan-Boltzman law $\rho_\gamma=\pi^2 T_{nl}^4/15$ results,
\begin{equation}\label{ratiorhos}
r(\eta_{nl},L)= \frac{\Delta\rho_B(\eta_{nl},L)}{\rho_\gamma} =
\frac{3 \times 2^{13} \; \pi^{\frac{11}{2}} }{\lambda^2} \;
\frac{N \; \alpha}{[ {\tilde\mu} \; \xi(\eta_{nl})]^4 \;
\sigma_R^2 \; \xi(\eta_{nl}) \; L}\;\frac{1}{(L\, T_R)^4} ~~;~~ L
\gg   \xi(\eta_{nl})\;.
\end{equation}
This result is the same as for the Minkowski space-time (see
eq.(7.53) in ref.\cite{magfiI}), except for a multiplicative
factor $ {\tilde\mu}^6 \; \xi^6(\eta_{nl}) \simeq 8 \;
\ln^3\left(\ln\frac{1}{\lambda}\right) $ and  the expression for
the correlation length in the radiation dominated universe
(\ref{corrlength}).

The factor $(LT_R)^{-4}$ is purely dimensional and is ultimately
the determinining factor for the strength of the generated
magnetic fields on a given scale. These combinations are
\emph{invariant} under cosmological expansion and are determined
by the ratio of the scales of interest today (galactic) to the
thermal wavelength (today) of the cosmic microwave background
radiation at the Wien peak. In particular $LT_R \sim 10^{25}$ for
$L \sim 1~\mbox{Mpc}$(today) [see eq.(\ref{LTR})].

It is clear that the production during this regime is extremely
small, due to the large values of $(LT)^4$ and of the ratio
$\sigma_R^2/\mu^2$. In order to obtain an estimate for the
amplitude of the seed magnetic field, we consider the following
set of parameters: $\lambda=10^{-2},\;\alpha= 10^{-2}, \; \mu=
 10^{14} \; \mbox{GeV}, \; T_R=10^{16} \; \mbox{GeV}$
(corresponding to a critical temperature $T_c= 10^{15}$ GeV). We
then obtain,
\begin{equation}
 r(L=1Mpc)\sim 10^{-157} \;.
\end{equation}
Therefore, the amplitude of the magnetic field generated during
the spinodal stage is completely negligible. This result is
similar to  the result  obtained in Minkowski space-time in
ref.\cite{magfiI} and is  expected on the basis of dimensional
analysis.

\subsection{Magnetogenesis from the scaling regime}

In the scaling regime $\eta >>\eta_{nl}$ the spectrum of the
magnetic field is given by (\ref{SButil}) with the mode functions
in the scaling regime given by (\ref{scaleform}).

The final expression for the leading contribution, given by
(\ref{SButil}) reveals a noteworthy aspect. As we have argued
above, the modes $k$ of astrophysical relevance today, were well
outside the horizon during the radiation dominated era between
reheating and the QCD phase transition. The mode functions
(\ref{SButil}) attain the largest amplitude at long times for
$x=q\eta \leq 2-3$, thus momenta in the polarization loop that are
within the horizon lead to generation of magnetic fields with
long-wavelengths well outside the horizon. This we believe, is an
important mechanism, loop corrections lead to a coupling between
modes inside the horizon with those outside. Thus in this manner,
causal fluctuations can actually lead to the generation of fields
with wavelengths much larger than the horizon.

Since $ k\eta \ll 1$ the power spectrum (\ref{SButil}) takes the
following form using the scaling mode functions (\ref{scaleform}),
\begin{equation}\label{SBesca}
S_B(\eta,k)=(1+2n_0)^2 \;\frac{\alpha N~k^2}{\pi~\sigma^2_R}\;
|A_0|^2 \int_0^{\infty} dq \; \int_{\eta_R}^{\eta} J_2(q\eta_1) \;
\eta_1\; d\eta_1 \int_{\eta_R}^{\eta} J_2(q\eta_2) \;\eta_2 \;
d\eta_2 \; I(q,k,\eta_1,\eta_2) \; ,
\end{equation}
where we set the exponentials equal to unity in eq.(\ref{SButil})
since $  k\eta \ll 1 $ and $ k \ll \sigma_R $ and
$$
I(q,k,\eta_1,\eta_2) \equiv \int_{-1}^{+1}dx \; \frac{1-x^2}{(q^2
+ k^2 - 2kqx)^2} \; J_2(\sqrt{q^2 + k^2 - 2kqx} \; \eta_1) \;
J_2(\sqrt{q^2 + k^2 - 2kqx} \; \eta_2) \; .
$$
Using the summation theorem\cite{gr}
$$
J_2(\sqrt{q^2 + k^2 - 2kqx}\eta) = \frac{4 (q^2 + k^2 - 2kqx)}{q^2
\, k^2 \, \eta^2}\sum_{l=0}^{\infty} (l+2)\;  J_{l+2}(q \, \eta)
\; J_{l+2}(k \, \eta) \; C_l^2(x)
$$
where the $ C_l^2(x) $ are Gegenbauer polynomials. For $  k\eta
\ll 1 $ the $ l = 0 $ terms dominate and we can use the small
argument behaviour of the Bessel functions  $ J_2(k \, \eta) =
\frac18 (k \, \eta)^2 [  1 + {\cal O}(k^2 \, \eta^2)] $. We
finally obtain,
\begin{equation}\label{Ifin}
I(q,k,\eta_1,\eta_2) = \frac{4}{3 \; q^4} \; J_2(q\eta_1) \;
J_2(q\eta_2) \left[ 1 + {\cal O}(k^2 \, \eta^2) \right] \; .
\end{equation}
Inserting eq.(\ref{Ifin}) into eq.(\ref{SBesca}) yields
\begin{equation}\label{SBes2}
S_B(\eta,k)=(1+2n_0)^2 \;\frac{\alpha N~k^2}{3 \;
\pi~\sigma^2_R}\; |A_0|^2 \int_0^{\infty} \frac{dq}{q^4} \left\{
\eta^2 \left[ J_2^2(q\eta) - J_1(q\eta) \; J_3(q\eta)\right] - (
\eta \rightarrow \eta_R ) \right\}^2 \; \left[ 1 + {\cal O}(k^2 \,
\eta^2) \right] \; ,
\end{equation}
where we used  the formula\cite{gr}
\begin{equation}
\int_{0}^{y} z \; J^2_2(\beta z)  \; dz=
\frac{y^2}{2}\left[J^2_2(\beta y)-J_1(\beta y) \; J_3(\beta y)
\right] \; .
\end{equation}
Since $ \eta \gg \eta_R $ we can neglect the terms with $ \eta_R $
 and we find, for $k\ll\eta^{-1} $
\begin{equation}\label{SBesf}
S_B(\eta,k)= {\cal D} \; \frac{\alpha N}{\lambda^2} \;
\frac{k^2}{\sigma^2_R}\; \mu^4 \; H_R^4 \; \eta^7\left[ 1 + {\cal
O}(k^2 \, \eta^2) \right]  \quad; \quad {\cal D}= 48.61\ldots
 \; .
\end{equation}
where we used eq.(\ref{A0^2}) and we computed numerically the
integral
$$
\int_0^\infty \frac{dx}{x^4}\;[J^2_2(x)-J_1(x)J_3(x)]^2 =
0.0005295\ldots
$$
This integral is dominated by the region $x  \geq 1$, namely, by
modes that are inside the horizon. From the estimate
(\ref{produ}), the corrections ${\cal O}\left(k^2
\xi^2_{diff}(\eta)\right)$ are truly negligible between reheating
and the QCD phase transition.

The dependence on the conformal time $\sim \eta^7$ is a direct
 consequence of the scaling form of the solution for the mode
 functions. The strong time dependence is a consequence of the causal
 relaxation of the Goldstone fields, a result of the phase ordering
 kinetics that entails that the size of the domains grow with the horizon.

There are several important features of the above result which are
noteworthy:

\begin{itemize}
\item{{\bf i:} The exponential associated with the diffusion
length cancels out, a reflection that the long time behavior of
the integrals above are dominated by the upper limit. Hence the
final result for the spectrum does not feature the exponential
suppression with the diffusion length. }

\item{{\bf ii:} The result for the spectrum only depends on the
initial amplitud $A_0$ and initial occupation number $n_0$ in the
combination $|A_0|^2(1+2n_0)$ which is constrained by the sum rule
(\ref{sumrule}). Hence the final spectrum is \emph{insensitive} to
the initial conditions on the mode functions or occupations, which
in principle carry information of the early history beginning from
the inflationary stage. This is a consequence of the scaling
solution being a fixed point of the dynamics of the scalar
field\cite{scaling,turok,durrer}. }

\item {{\bf iii:} A noteworthy result is that superhorizon
magnetic fields are generated by the non-equilibrium dynamics of
modes inside but near the Hubble radius. This is a consequence of
the polarization loop, wherein the propagators correspond to
momenta $q$ and $|{\vec q}+{\vec k}|$. The momenta $k$
corresponding the wavevector (scale) of the magnetic field is such
that the wavelength is larger than the Hubble radius, but the
momenta $q$ corresponding to the charged scalar field fluctuations
are inside the horizon. The correlation length of the charged
scalar field is of the order of the Hubble radius. Thus acausal,
superhorizon magnetic fields are generated by \emph{loop} effects.
Furthermore, the processes that yield the leading contribution are
pair \emph{production}, pair annihilation and low energy
bremsstrahlung. These processes are not allowed in equilibrium
because of energy momentum conservation, but are allowed in a
rapidly expanding cosmology and strongly out of equilibrium as is
the case under study.}

\item{{\bf iv:} The out of equilibrium dynamics for very long
superhorizon wavelengths cannot be captured with a kinetic
description as advocated in ref.~\cite{giovashapo}, since kinetic
descriptions rely on slow time variations of the distributions.
Clearly such description will not be justified during the early
and intermediate stages of the evolution after the phase
transition because of the rapid (exponential) time evolution of
the mode functions and therefore the distribution functions. The
correct scaling dynamics cannot be captured within the kinetic
description either since it describes the evolution of coherent
domains correlated on scales comparable to the Hubble radius. Thus
the formulation presented here based on the Schwinger-Dyson
equations provides the necessary calculational scheme to describe
reliably from first principles the generation of magnetic fields
during and after non-equilibrium phase transitions. }

\end{itemize}

In order to reveal the enhancement during the scaling regime in a
more transparent manner,  it is convenient to use the relations
(\ref{H-T}),  (\ref{eta.T}), (\ref{crittemp}) and the explicit
expression for the conductivity (\ref{sigmacond}) in the form
\begin{equation}\label{conductivity}
\sigma_R=c[\alpha,N]\; \frac {N~T_R}\alpha \quad , \quad
c[\alpha,N] \equiv \frac{\mathcal{C}}{\ln[\frac{1}{\alpha \, N}]}
\sim {\cal O}(1).
\end{equation}
Then, the  ratio $r(L,\eta)$ for $ L \gg \eta $ is given by
\begin{equation}\label{rreges}
r(\eta,L) = \frac{240\pi~ {\cal D} ~\alpha^3}{N~c^2[\alpha,N] \;
[LT_R]^5} \; \left(\frac{\mu}{\sqrt{\lambda}T(\eta)}\right)^4 \;
\left(\frac{M_*}{T(\eta)}\right)^3 ~.
\end{equation}
where ${\cal D}$ is given in eq. (\ref{SBesf}).  We note that in
the final result (\ref{rreges}) there is no dependence on the
reheating temperature  but only on the scale of symmetry breaking
$\mu$, the temperature at the time $\eta$ and the scalar and gauge
couplings. This is expected since the non-equilibrium processes
begin in earnest after the phase transition,  local thermal
equilibrium prevailed between the time of reheating and the phase
transition.

The dependence on the scalar self coupling $\propto 1/\lambda^2$
is a hallmark of the non-perturbative nature of the growth of
unstable modes and spinodal decomposition, it is ubiquitous in the
non-equilibrium dynamics of phase
transitions\cite{scaling,nuesfrw,nuestros}.

Large scale magnetogenesis is more efficient for large symmetry
breaking scale $\mu$, since the larger the symmetry breaking
scale, the longer  lasts the scaling stage.

Consider for instance the case in which the symmetry breaking
scale $\mu \sim 10^{13}~\mbox{Gev}$ and $\lambda \sim \alpha \sim
10^{-2}$, corresponding to a critical temperature of order of  a
GUT scale $T_c\sim 10^{15} \mbox{GeV}$ and suppose that the
scaling regime lasts until the electroweak phase transition scale,
i.e. $\eta$ is such that $T(\eta)=T_{EW}\sim 10^2~\mbox{GeV}$.
Then the factor
$$
\left(\frac{\mu}{\sqrt{\lambda}T_{EW}}\right)^4
\left(\frac{M_*}{T_{EW}}\right)^3\sim 10^{100}
$$
compensates for the factor $(LT_R)^{-5}$. Taking $N$ and $g_*$ of
the order of $10$ (these values are taken as representative and
they can be changed simply in the final expressions) we can write
the expression for the ratio as
\begin{equation}\label{r-EW}
r(T(\eta),L) \simeq 10^{-34} \left(
\frac{L}{1~\mbox{Mpc}}\right)^{-5} \left(\frac{T_{EW}}{T(\eta)}
\right)^7 \; .
\end{equation}
Therefore \begin{equation}\label{finrat}r(T(\eta),L) \sim \left\{
\begin{array}{l}
  10^{-34}~\mbox{at\,the\,EW\,transition} \\
  10^{-14}~\mbox{at\,the\,QCD\,transition}\; .
\end{array}\right.
\end{equation}
Thus if the scaling regime lasts until a time between the EW and
the QCD phase transitions the amplitude of the large scale
magnetic fields is within the range necessary to be amplified by
some dynamo models. The amplitude of the seed magnetic field
 is strongly dependent on the duration of the scaling
regime. We have only focused on a scaling regime terminating
either at the EW or QCD phase transition since there will surely
be new phenomena associated with these that must be included in
the dynamics of magnetogenesis.

\subsection{Discussion}

\begin{itemize}

\item{{\bf Validity of the approximations:} There are two main
approximations that were used to obtain the results quoted above,
i) the long-wavelength approximation $k\eta \ll 1$ and ii) the
weak coupling approximation. We now provide an estimate of the
reliability of both these approximations to establish the limit of
validity of our results.

\bigskip

{\bf i): Long-wavelength approximation:} In order to reach our
final result for the rate $r(L,\eta)$  we have explicitly used a
series of approximations which are valid for long wavelengths but
whose validity must be checked before we reach any conclusion
regarding the spectrum at \emph{small} scales. In particular we
must address the limits of applicability of the result
(\ref{rreges}). This result has been obtained by integrating the
magnetic spectrum on scales $0<k<k_{max}$ with
$k_{max}=2\pi/L_{min}$; the formula for the magnetic spectrum was
valid in the limit
\begin{equation}
k_{max} \; \eta_{max}\ll1\;.
\end{equation}
In order to provide an estimate may take for $\eta_{max}$ to be
the (conformal) time at which the EW phase transition occurs,
namely $\eta_{EW}\sim 1  \; \mbox{GeV}^{-1}$. As discussed in the
introduction, we are considering  a situation in which the
magnetic field is considered as a perturbation of a pre-existing
thermal blackbody background. For consistency this requires that,
\begin{equation}
r(L_{min},\eta_{WE})\ll1\;.
\end{equation} This
relation translates in a condition
\begin{equation}
k_{max} \; \eta_{max}\ll\left[\frac{C
N\alpha^3}{24^2c^2}\left(\frac{T_c}{T_{EW}}
\right)^4\left(\frac{T_{EW}}{M_*}\right)^2\right]^{-1/5}\;.
\end{equation}
For $T_c\sim 10^{15} \mbox{GeV}$ this gives $k_{max} \;
\eta_{max}\ll 0.0176$ which in turns translate into
\begin{equation} L\gg L_{min}\sim70 \mbox{ fm}\;.
\end{equation} However, this is the comoving length normalized at
the reheating time. In order to convert to the present time, we
have to take in account the redshift
\begin{equation}
z_R=\frac{T_R}{T_0}\simeq 4\times 10^{28}~~\mbox{for}~T_R\sim
10^{15}\mbox{Gev} \;;
\end{equation}
this gives
\begin{equation}
\left. L_{min}\right|_{today}\sim 0.1 \mbox{ pc}\;.
\end{equation}
 Thus, the approximations
invoked are reliable to estimate the amplitude of primordial seeds
on galactic scales or larger, today.

\bigskip

{\bf ii): Weak electromagnetic coupling:} In order to study the
amplitude for much smaller scales the calculations must be done
without the long wavelength approximations  invoked above. In this
case we must expect a breakdown of perturbation theory and we
cannot give a reliable estimate in the present framework.
Furthermore, for scales well inside the Hubble radius,
microphysical processes \emph{not included} in our approximations,
such as scattering between charged fields and between charged  and
gauge fields must be included. These processes will restore
equilibrium between the different fields, if there is a
substantial transfer of power from the charged fluctuations to the
radiation field, this may lead to a change in the equation of
state and the full backreaction on the metric must be included. At
longer time scales the effects of the backreaction of the gauge
fields on the dynamics of the scalar field, as well as the
non-equilibrium contributions to equation of state and the
Friedmann equations must be included self-consistently. }

\item{{\bf Generation on short distance scales:  } For scales well
inside the horizon during the scaling regime, namely $q\eta \gg
1$, we must  account for causal microscopic processes that tend to
equilibrate the electromagnetic fields generated by the
non-equilibrium processes. In order to understand these processes
we must look at the kinetics of equilibration. The mode functions
for wavevectors well inside the horizon are Minkowski-like, of the
form
$$
f_q(\eta)= \frac{\alpha_q}{\sqrt{q}} \;
e^{-iq\eta}+\frac{\beta_q}{\sqrt{q}} \; e^{iq\eta} \;  ,
$$
where the coefficients $\alpha_q,\beta_q$ must be determined from
a full numerical evolution. However the constancy of the Wronskian
entails that
$$
|\alpha_q|^2-|\beta_q|^2 =1
$$
which suggests the identification $|\alpha_q| \equiv
1+\mathcal{N}_q\,;\,|\beta_q|\equiv \mathcal{N}_q$,
$\mathcal{N}_q$ is the number of (asymptotic) quanta created
during the time evolution. This form of the asymptotic mode
functions leads to the equipartition between the electric and
magnetic field generation, since spatial and time derivatives are
the same. In turn this entails that we can understand the
generation of electric and magnetic fields by obtaining a
\emph{kinetic} equation for the number  of photons. Such kinetic
equation must necessarily be of the form
$$
\frac{d N_k(\eta)}{d\eta} =
[1+N_k(\eta)]\Gamma^>_k(\eta)-N_k(\eta)\Gamma^<_k(\eta)
$$
which displays the familiar gain minus loss contributions in terms
of the forward and inverse rates. Eventually a steady state will
be reached which will describe a stationary distribution of
photons. The computation of the forward [$\Gamma^>_k(\eta)$] and
inverse [$\Gamma^<_k(\eta)$] require a \emph{detailed} knowledge
of the distribution $\mathcal{N}_q$\cite{Boyanovsky:1999jh} since
these generalized rates are functionals of these occupation
numbers. Clearly such computation lies beyond the scope of this
article and is a task that we will undertake elsewhere. However,
the kinetic equations above will tend to an equilibrated state of
local thermodynamic equilibrium. }

\item{{\bf Effect on the LSS:} It is important to estimate the
effect of the magnetic field on scales corresponding to those of
the last scattering surface, which today are $L_{LSS}\sim
100~\mbox{Mpc}$. From eq. (\ref{finrat}) we see that at the
electroweak temperature $r(T_{EW},L_{LSS}) \sim 10^{-44}$, taking
the fourth root we can provide an estimate of the temperature
fluctuation induced by the primordial magnetic field $\left.\delta
T/T \right|_{LSS} \sim [r(T_{EW},L_{LSS})]^{\frac{1}{4}} \sim
10^{-11}$ which is negligible compared to the CMB temperature
fluctuation at this scale $\sim 10^{-5}$. On the other hand, a
similar estimate at the time of the QCD phase transition gives
$\left.\delta T/T \right|_{LSS} \sim 10^{-6}$ which is marginally
compatible with the current observations. Thus the reliability of
the approximation of weak gauge coupling combined with the effects
on the temperature anisotropy at the last scattering surface seem
to lead us to conclude that \emph{if} a phase transition during a
radiation dominated era occurs near the GUT scale and results in a
scaling stage, our results for primordial magnetogenesis will be
reliable down to the scale of electroweak symmetry breaking. }

\end{itemize}

\section{Conclusions}

We review here large scale primordial magnetogenesis
during a phase transition in the radiation dominated era after
reheating in a model of $N$-charged scalars coupled to an abelian
gauge field\cite{magfiI}. The spectrum of the magnetic field generated during
the non-equilibrium evolution was computed using the formulation
recently introduced in ref.\cite{magfiI}. The dissipative effects
of the conductivity are included by separating the contribution
from hard modes (with momenta of order $T$) to the polarization
tensor of the gauge fields. These modes  are always in local
thermodynamic equilibrium. The non-perturbative, non-equilibrium
dynamics of the scalar field after the phase transition was
studied in the large $N$ limit. The dynamics after the phase
transition features two distinct stages: an early and intermediate
time, spinodal stage, which is dominated by the growth of
long-wavelength fluctuations, followed by a scaling regime during
which the scalar field becomes correlated over horizon-sized
domains. During both regimes, strong non-equilibrium fluctuations
lead to large current-current correlation functions which entail
the generation of magnetic fields. The scaling regime is the most
effective for primordial magnetogenesis since this stage lasts the
longest. During this stage magnetic fields with superhorizon
wavelengths are generated via \emph{loop} effects, the scalar
field momenta in the polarization loop corresponds to wavelengths
of the order of or shorter than the horizon. Thus causal scalar
field fluctuations lead to the generation of magnetic fields on
superhorizon scales. In particular the processes that yield the
leading contribution to magnetic field generation are i): pair
production, ii) pair annihilation and iii) low energy
bremsstrahlung, which cannot occur in equilibrium because of
energy momentum conservation, but do occur out of equilibrium
because of the rapid time evolution due to the expansion and the
evolution of the scalar field out of equilibrium.

 The generation of magnetic field is hindered
by the large conductivity of the plasma and equipartition between
electric and magnetic fields does not hold. The spectrum of the
primordial magnetic field is insensitive to the magnetic diffusion
length which is sub-horizon during the radiation era.

Our final result for the spectrum generated during the scaling
regime is given by eq.(\ref{SBesf}). The  ratio of the energy
density of the magnetic fields on scales larger than $L$ to the
energy density in the cosmic background radiation
$r(L,\eta)=\rho_B(L,\eta)/\rho_{cmb}(L,\eta)$ is given by equation
(\ref{rreges}). For values of $N$, and the gauge coupling
consistent with particle physics models we find that
\begin{equation}\label{r-EW2}
r(T(\eta),L) \simeq 10^{-34} \left(
\frac{L}{1~\mbox{Mpc}}\right)^{-5} \left(\frac{T_{EW}}{T(\eta)}
\right)^7\; .
\end{equation}
Therefore,
\begin{equation}\label{finrat2}r(T(\eta),L) \sim \left\{
\begin{array}{l}
  10^{-34}~\mbox{at\,the\,EW\,transition} \\
  10^{-14}~\mbox{at\,the\,QCD\,transition} \; .
\end{array}\right.
\end{equation}
Therefore, the large scale primordial magnetic fields generated
during the scaling stage after a phase transition may be a
plausible mechanism to generate primordial magnetic fields which
will  be further amplified by the collapse of protogalaxies and by
astrophysical dynamos.

Probably a phase transition at a temperature much larger than the
electroweak leading to a scaling regime lasting until the QCD
phase transition is ruled out by the temperature inhomogeneities
at the last scattering surface. Furthermore the generation of
electromagnetic fields on sub-horizon scales requires a full
kinetic equation that incorporates the microscopic causal
processes that lead to thermalization, the study of these is
beyond the scope of this article.

{\bf Magnetogenesis \emph{after} the QCD phase transition?} The
model that we studied here is assumed to describe the robust
features from the non-equilibrium dynamics of a charged sector
coupled to a (hyper) charge gauge field. GUT's or SUSY theories
may provide the corresponding framework.

However we now argue that precisely the model studied here can
actually describe the non-equilibrium dynamics \emph{after} the
QCD phase transition(s).  After hadronization and chiral symmetry
breaking most of the hadrons produced will be pions, at least this
is the experimental situation in ultrarelativistic heavy ion
collisions. Neglecting the charge form factor (which is justified
for momenta much smaller than the $\rho$ meson mass $m_\rho \sim
770 ~\mbox{Mev}$) the charged pions couple to the electromagnetic
field with minimal coupling. The chiral transition is conjectured
to be in the same universality class as the $O(4)$ linear sigma
model\cite{wilraja}. Thus the model presented in this article, is
the \emph{low energy effective field theory} for the triplet of
pions, two charged and one neutral.

Thus we conjecture that the study in this article can be applied
to the study of the generation of magnetic fields by
long-wavelength pions. Therefore the analysis of this article can
apply to magnetogenesis during the \emph{chiral} phase transition
in QCD. While the charged pions couple to electromagnetism via the
minimal coupling in the long-wavelength limit, the neutral pion
couples to the electromagnetic field through the chiral anomaly
$\pi^0 \rightarrow 2\gamma$ and such process will also produce
magnetic fields.

Since the ratio of the relaxation rate via the strong interactions
to the expansion rate

\begin{equation}
\frac{\Gamma_{QCD}}{H(\eta_{QCD})} \sim 10^{28}
\end{equation}

\noindent then if the chiral phase transition is second order, it
will occur in LTE. However if it is first order as suggested by
the fact that the pions are massive, then non-equilibrium effects
may arise and primordial magnetic fields may be generated by
charged pion bremsstrahlung or pair annihilation or by neutral
pion decay. These possibilities can be studied with the linear
sigma model effective field theory.

{\bf Acknowledgements:}  We would like to thank Norma Sanchez for 
organizing this wonderful and very stimulating IX Chalonge School of
Astrofundamental Physics.
The authors thank M. Giovannini, for
useful discussions. D. B. and M. S. thank N.S.F. for support
through grants PHY-9988720 and NSF-INT-9815064.

\end{document}